\documentclass[lettersize,journal]{IEEEtran}
\usepackage[utf8]{inputenc} 
\usepackage[T1]{fontenc}
\usepackage{url}
\usepackage{ifthen}
\usepackage[cmex10]{amsmath}
\usepackage{array}

\usepackage{subfigure}

\usepackage{enumerate}
\usepackage{amssymb}
\usepackage{amsmath} 

\usepackage{booktabs}
\usepackage{multirow}

\usepackage{algorithm}
\usepackage{algorithmicx}
\usepackage{algpseudocode} 
\usepackage{subfig}
\usepackage{cite}

\usepackage{threeparttable}

\usepackage{mathrsfs}
\usepackage{bm}
\usepackage{tikz}
\usetikzlibrary{arrows}
\usepackage{graphicx,booktabs,multirow}
\usepackage{epstopdf}
\usepackage{multirow}
\usepackage{tabularx} 
\usepackage{booktabs}
\usepackage[font={small}]{caption}

\usepackage{pifont}

\definecolor{colorhkust}{RGB}{20,43,140}
\definecolor{colortsinghua}{RGB}{116,52,129}
\definecolor{color1}{RGB}{128,0,0}

\usepackage{amsthm}
\usepackage{cite}

\newtheorem{remark}{Remark}

\theoremstyle{definition}

\theoremstyle{remark}

\begin{document}

        
        
        
        
        
        
        
        \title{Task-Oriented Communication for Edge Video Analytics}
        
        
        

      \author{Jiawei~Shao,~\IEEEmembership{Graduate Student Member,~IEEE,}
      Xinjie~Zhang,~\IEEEmembership{Graduate Student Member,~IEEE,}
        and~Jun~Zhang,~\IEEEmembership{Fellow,~IEEE}
      	\thanks{The authors are with the Department of Electronic and Computer Engineering, Hong Kong University of Science and Technology, Hong Kong (E-mail: \{jiawei.shao,xinjie.zhang\}@connect.ust.hk, eejzhang@ust.hk). The corresponding author is J. Zhang. This work was supported by the General Research Fund (Project No. 15207220) from the Hong Kong Research Grants Council.}

}
\maketitle
\begin{abstract}

With the development of artificial intelligence (AI) techniques and the increasing popularity of camera-equipped devices, many edge video analytics applications are emerging, calling for the deployment of computation-intensive AI models at the network edge.
Edge inference is a promising solution to move computation-intensive workloads from low-end devices to a powerful edge server for video analytics, but device-server communications will remain a bottleneck due to limited bandwidth.
This paper proposes a task-oriented communication framework for edge video analytics, where multiple devices collect the visual sensory data and transmit the informative features to an edge server for processing.
To enable low-latency inference, this framework removes video redundancy in spatial and temporal domains and transmits minimal information that is essential for the downstream task, rather than reconstructing the videos on the edge server.
Specifically, it extracts compact task-relevant features based on the deterministic information bottleneck (IB) principle, which characterizes a tradeoff between the informativeness of the features and the communication cost.
As the features of consecutive frames are temporally correlated, we propose a temporal entropy model (TEM) to reduce the bitrate by taking the previous features as side information in feature encoding.
To further improve the inference performance, we build a spatial-temporal fusion module on the server to integrate features of the current and previous frames for joint inference.
Extensive experiments on video analytics tasks evidence that the proposed framework effectively encodes task-relevant information of video data and achieves a better rate-performance tradeoff than existing methods.



\end{abstract}


%
\IEEEpeerreviewmaketitle

\section{Introduction}

Video analytics has grown substantially from practical needs in the past decade, driven by the revival of artificial intelligence (AI) and the advancement of deep learning.
AI-enabled video analytics applications are being widely deployed across different fields such as smart transportation \cite{chang2020ai_smart_transportation}, security surveillance systems \cite{othman2017new_surveillance}, and online healthcare \cite{gao2018computer_healthcare}.
The main goal is to automatically recognize temporal-spatial events in videos and gain valuable insights.
At the core of these applications lie powerful deep learning models performing tasks including image classification, object detection, and semantic segmentation.

As edge devices equipped with camera sensors (e.g., drones, mobile phones, and monitoring cameras) are becoming increasingly widespread, the industry has shown strong interest in deploying video analytics at the network edge to support ubiquitous AI-enabled services, which gives rise to a new research area named \emph{edge video analytics} \cite{ananthanarayanan2017real_video_analytics}.
As the deep learning models demand intensive computation, leading to a long delay when running on resource-constrained edge devices, typical edge systems forward all the input data from the edge devices to a proximate edge server (e.g., a base station) to perform inference.
However, when devices are connected wirelessly, such as wearable devices and autonomous cars, the available uplink bandwidth is very limited and the communication channel is highly dynamic.
The enormous volume of the collected videos leads to prohibitive communication overhead if directly transmitting them to the edge server.
Meanwhile, as many video analytics applications have stringent response time requirements, the high communication latency hinders user experience.
This tension between the communication overhead and the latency requirement brings the main design challenge of edge video analytics systems.

To reduce the bandwidth consumption and response time in edge video analytics, a \emph{cooperative edge inference} framework stands out as a promising solution \cite{zhou2019edge_edge_intell}, \cite{ShaoTWC}.
With multiple devices at the network edge that capture video data, each device first extracts compact features from input frames that preserve the task-relevant information and then forwards them to be processed by an edge server. 
This exemplifies a recent shift in communication system design to support intelligence applications at the edge, named \emph{task-oriented communication} \cite{ShaoTWC,shao2021learning,hanna2020distributed,singhal2020communication,choi2019context,Multi-Robot_Collaborative_Percep_GNN,xie2022robust_IB}, which optimizes communication strategies for the downstream task, rather than for recovering data at the server as in traditional \emph{data-oriented communication}.
Many existing works have led to efficient communication strategies for edge inference.
For example, the VFE method proposed in \cite{shao2021learning} formalized a tradeoff between the communication rate and accuracy based on the information bottleneck (IB) principle \cite{tishby2000informationIB} for single-device edge inference.
Later, the authors of \cite{ShaoTWC} considered the task-oriented communication framework for multi-view computer vision tasks in edge inference.
They developed a distributed feature encoding method to reduce the communication volumes, which squeezes the spatial redundancy among the extracted features of multi-view images.

While cooperative edge inference with task-oriented communication has enabled low-latency edge applications in previous works, directly applying these methods to edge video analytics may lose their effectiveness due to two main reasons.
First, identifying the task-relevant features from videos requires a high-complexity extractor, which needs to process a sequence of frames by either recurrent neural networks \cite{sundermeyer2012lstm_LSTM} or attention-based transformer blocks \cite{vaswani2017attention}.
These two AI models demand much higher computation and memory resources than convolutional neural networks \cite{ResNet} adopted in previous studies on image-related tasks \cite{shao2021learning, ShaoTWC}.
Second, previous works neglected the copious temporal information during the feature encoding process, which resulted in high redundancy for video data.
In many applications such as pedestrian tracking \cite{ristani2018features_pedestrian_tracking} and vehicle re-identification \cite{liu2016deep_vehicle_reid}, it is intuitive that motion estimation \cite{farneback2003two_motion_estimation} can help to exploit the temporal correlation among adjacent frames to substantially reduce the communication cost since the backgrounds of consecutive frames are visually similar.
This observation motivates us to develop a task-oriented communication framework with a temporal entropy model (TOCOM-TEM) to tackle the aforementioned problems and accelerate video analytics at the network edge.
Particularly, we leverage the \emph{deterministic information bottleneck} principle \cite{deterministic_ib} to extract the task-relevant feature from each frame independently and develop a spatial-temporal fusion module at the server to jointly leverage the current received features as well as the previous features to perform inference.
The temporal entropy model helps to reduce the encoded feature size by using the previously transmitted features as side information.

\subsection{Related Works}

The recent investigation of task-oriented communication \cite{strinati20216g_goal_oriented,ShaoTWC,zhu2020toward,shao2020communication,pappas2021goal,shlezinger2021deep,merluzzi2021wireless}, as well as similar ideas such as semantic communication \cite{xie2021deep_semantic_NLP}, \cite{bourtsoulatze2019deep}, has motivated a paradigm shift in communication system design.
The conventional \emph{data-oriented communication} aims to guarantee the reliable transmission of every single bit of the raw data, while oblivious to the message semantics or the downstream task.
For edge video analytics, it will incur excessive delay if the high-volume video data is directly transmitted with the limited communication resources at the network edge.
The task-oriented communication principle is promising to alleviate this challenge.
Specifically, it discards information of the input frames (e.g., image noise and background) that is irrelevant to the given task, and transmits only the task-relevant feature (e.g., region of interest) that could influence the inference result.
In this way, it has the potential to achieve orders of magnitude reduction in the communication overhead, thus significantly reducing the transmission latency.

There have been several recent studies taking advantage of deep learning to develop effective feature extraction and encoding methods following the task-oriented design principle \cite{ShaoTWC,shao2021learning,iccshao,kang2022task_oriented_ref1,dubois2021lossy,li2023task_hongru}.
In particular, an end-to-end architecture for image classification was proposed in \cite{bottlenet}, which leverages a JPEG compressor to encode the internal features for the downstream task.
The proposed method was directly trained with the loss function specified by the task and did not take the feature recovery quality into consideration.
The same idea was utilized in other intelligence edge applications including image retrieval \cite{jankowski2020wireless_Jankowski}, point cloud processing \cite{shao2020branchy}, and semantic communication \cite{xie2021deep_semantic_NLP}.
Besides, the authors of \cite{shao2021learning} formalized a tradeoff between the communication rate and the inference performance in task-oriented communication based on the information bottleneck (IB) principle \cite{tishby2000informationIB}.
Notably, the IB principle provides the insight that a good feature should be sufficient for the inference task while retaining minimal information from the input data. This is consistent with the design objective in cooperative edge inference.
Meanwhile, the IB principle is also adopted in \cite{strinati20216g_goal_oriented} to identify the relevant semantic information to accomplish a specific goal.
The most recent work \cite{ShaoTWC} extended this task-oriented communication framework to multi-device edge inference systems, which developed a distributed feature coding method to squeeze the spatial redundancy among the extracted features of multiple devices.

Although the aforementioned works proposed several task-oriented communication strategies to reduce the communication overhead at the network edge, they mainly considered simple tasks with a single instance as the input data (e.g., image classification).
Directly applying them to tasks with sequential data (e.g., edge video analytics systems) is not efficient.
This is because the methods designed for single-frame processing cannot exploit the temporal correlation among a sequence of frames. 
Traditional video compression algorithms \cite{wiegand2003overview_AVC}, \cite{sullivan2012overview_HECV} adopt a predictive coding architecture to achieve a high compression ratio.
Particularly, these methods apply a motion-compensation loop at the encoder to explicitly investigate the inter-frame redundancy.
Recently, deep learning-based video compression methods have been developed \cite{lu2019_dvc, liu2020conditional_conditional_entropy_model,tung2022deepwive,liu2020deep,ma2019image,agustsson2020scale,hu2021fvc,li2021deep}, which enable end-to-end optimization to further improve the compression performance.
However, both the traditional and learning-based video codecs require a heavy encoder to exploit the statistical redundancy in the \emph{pixel domain}, aiming at recovering the video with high resolution under a bitrate constraint.  
This kind of data-oriented objective does not fit with video analytics applications at the edge, which motivates us to develop a task-oriented communication framework to exploit the temporal-spatial correlation of videos in the \emph{feature domain}.
Meanwhile, we will also take inspiration from deep learning-based compression methods.

\subsection{Contributions}
In this paper, we develop a task-oriented communication framework for edge video analytics, which effectively extracts task-relevant features and reduces both the spatial and temporal redundancy in the feature domain.
The major contributions are summarized as follows:
\begin{itemize}
\item This work formulates three design problems of task-oriented communication for edge video analytics, namely, task-relevant feature extraction, feature encoding, and joint inference. 
We propose a task-oriented communication (TOCOM) framework with a temporal entropy model (TEM) for edge video analytics, which is named TOCOM-TEM.

\item First, we leverage the deterministic information bottleneck to design the local feature extractor, which aims at maximizing the mutual information between the extracted feature and the desired output while minimizing the entropy of the feature. 
Thus, it addresses the objective of reducing communication overhead by discarding task-irrelevant information.
As the mutual information and entropy terms in the deterministic information bottleneck are computationally prohibitive for high-dimensional data, we resort to variational approximations to devise tractable upper bounds.
\item Second, to further compress the size of the extracted feature and reduce the joint bitrate, we perform entropy coding to encode the current feature by adopting the previous feature as side information.
We proposed a temporal entropy model to exploit the dependence among features by estimating the conditional probability distribution of the current feature given the previous ones.
\item Third, to better exploit both spatial and temporal correlation of video data collected from multiple devices, we construct a spatial-temporal fusion module at the server to integrate the current received features and the previous features to predict the desired output.
\item The effectiveness of the proposed framework is validated in two edge video analytics tasks, i.e., a multi-camera pedestrian occupancy prediction task and a multi-camera object detection task. 
Extensive simulation results show that our TOCOM-TEM method achieves a substantial reduction in the communication overhead compared with the image compression and video coding methods.
Our study demonstrates that the task-oriented design principle is promising to enable low-latency edge video analytics applications.
\end{itemize}

\subsection{Organization}

The rest of this paper is organized as follows. Section \ref{Sec:System Model and Problem Formulation} introduces the system model and formulates the design problems of task-oriented communication for edge video analytics, and Section \ref{Sec:Task-Orientend Communication with Temporal Entropy Coding} proposes the TOCOM-TEM method.
In Section \ref{Sec:Experiment}, we provide extensive simulation results to evaluate the performance of the proposed methods. Finally, Section \ref{Sec:Conclusion} concludes the paper.

\subsection{General Notations}

Throughout this paper, upper-case letters (e.g. $X,Y$) and lower-case letters (e.g. $\bm{x,y}$) stand for random variables and their realizations, respectively. 
We abbreviate a sequence $\{X_{1}, \ldots, X_{N}\}$ of $N$ random variables by $X_{1:N}$ and similarly use $X^{(1:K)}$ to represent the set of random variables $\{ X^{(1)},\ldots, X^{(K)} \}$.
The entropy of $Y$, the conditional entropy of $Y$ given $X$, and the joint entropy of $X,Y$ are denoted as $H(Y)$, $H(Y|X)$, and $H(X,Y)$, respectively. 
The mutual information between $X$ and $Y$ is represented as $I(X,Y)$.
The Kullback-Leibler (KL) divergence between two probability distributions $p(\bm{x})$ and $q(\bm{x})$ is denoted as $D_{\mathrm{KL}}\left(p||q\right)$, and the cross entropy between $p(\bm{x})$ and $q(\bm{x})$ is represented as $H(p,q)$.
The statistical expectation of $X$ is denoted as $\mathbb{E}\left(X\right)$.
We further denote the Gaussian distribution with vector ${\mu}$ and variance ${\sigma}$ as $\mathcal{N}\left({\mu},{\sigma}\right)$
The notation $\mathcal{U}(a,b)$ represents the uniform distribution with the range $[a,b]$.
The convolutional operation is denoted as $*$, and the quantization function is $\bm{\hat{x}} = \left\lfloor \bm{x} \right\rceil$.

\begin{figure*}[t]
    \centering
    \includegraphics[width=0.9\linewidth]{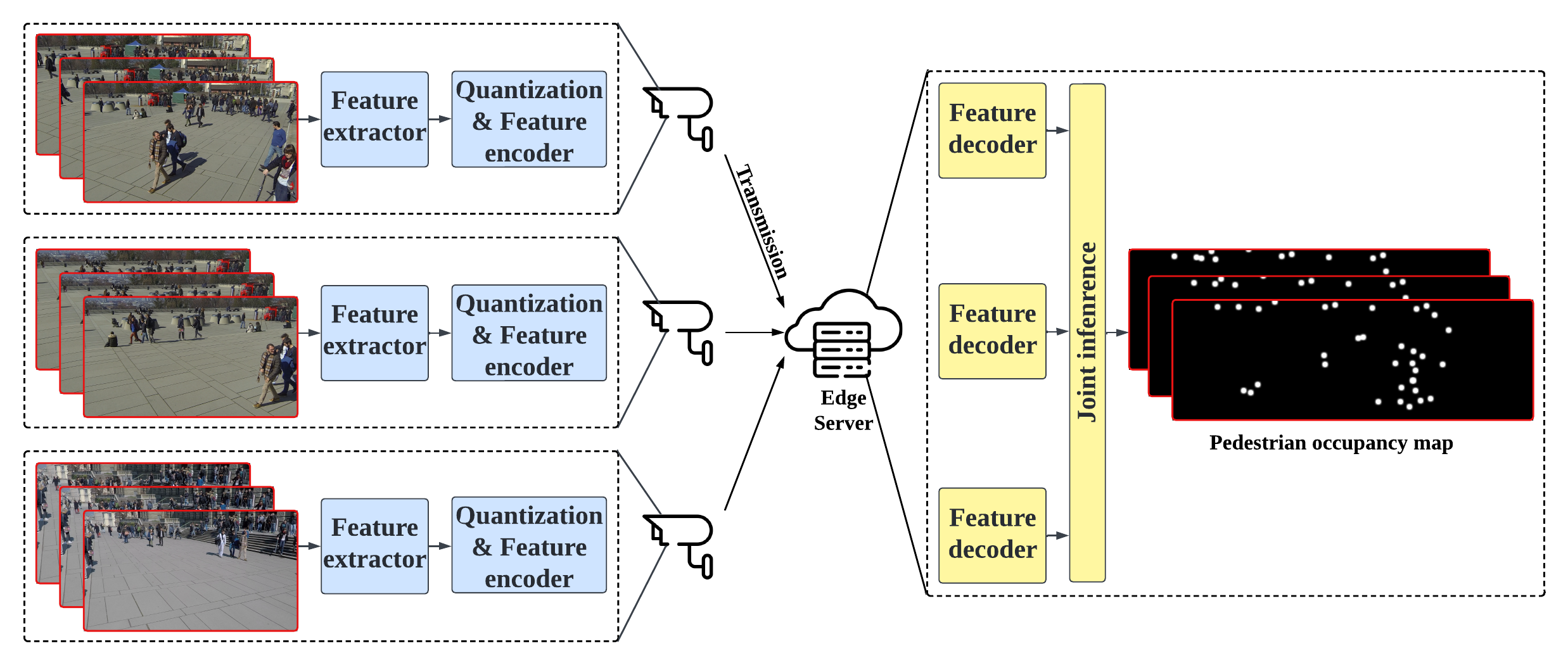}
    \caption{An example of edge video analytics for pedestrian occupancy prediction.
    The edge server takes consecutive video frames from multiple cameras as inputs to estimate the position of pedestrians.
    }
    \label{fig:sample_pedestrian_detection}
\end{figure*}

\begin{figure*}
    \centering
    \includegraphics[width = 0.95\linewidth]{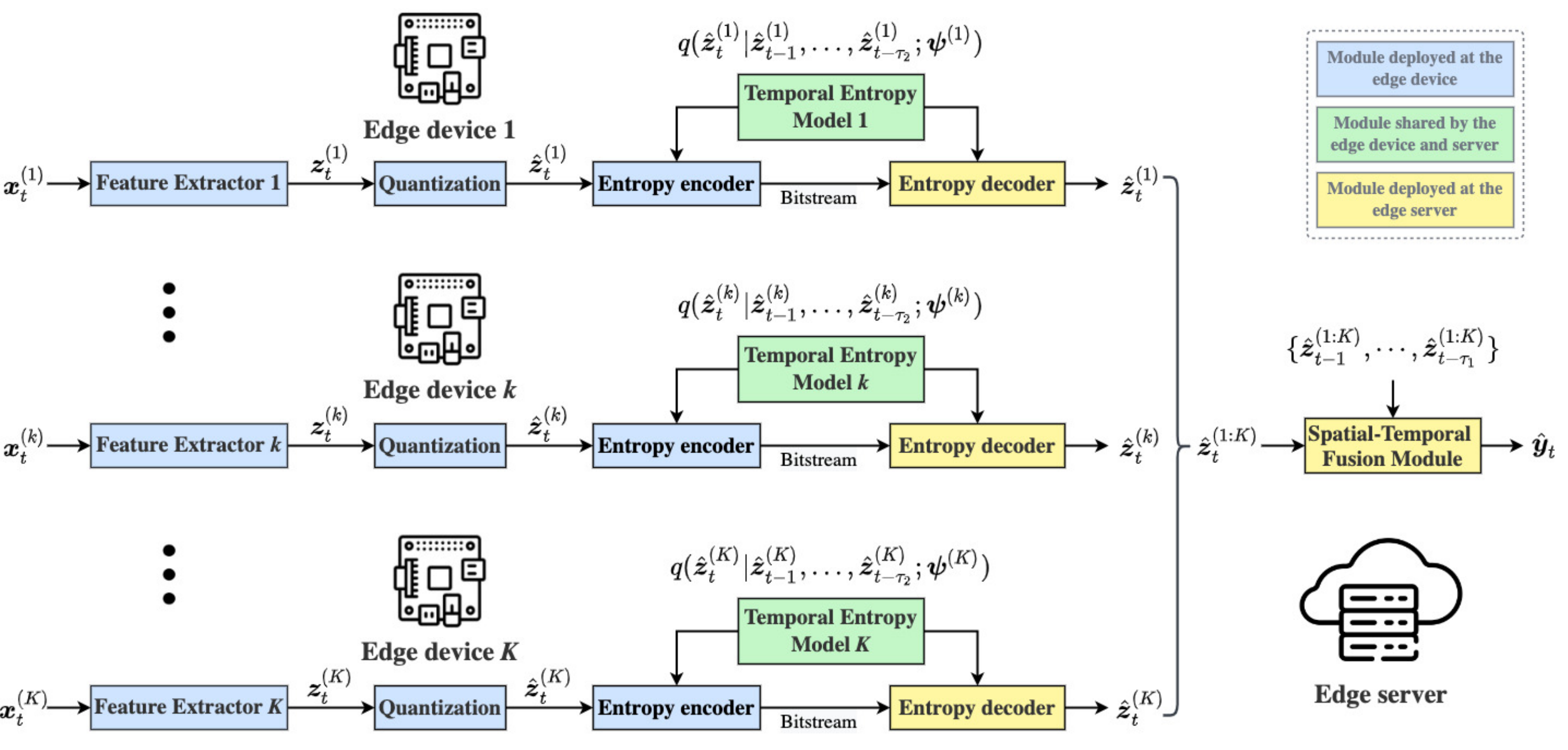}
    \caption{Probabilistic modeling for edge video analytics systems. At each time step $t$, edge devices extract the task-relevant features $\bm{z}_{t}^{(k)}$ from their inputs $\bm{x}_{t}^{(k)}$, adopt the temporal entropy model to encode the quantized feature $\hat{\bm{z}}_{t}^{(k)}$, and forward them to an edge server for further processing.
    A spatial-temporal fusion module is deployed at the edge server that jointly leverages the current received features $\hat{\bm{z}}_{t}^{(1:K)}$ as well as the previous features $\{ \hat{\bm{z}}_{t-1}^{(1:K)},\ldots, \hat{\bm{z}}_{t-\tau_{1}}^{(1:K)} \}$ to predict the target variable $\hat{\bm{y}}_{t}$.}

    \label{fig:edge_video_analytics_framework}
\end{figure*}

\section{System Model and Problem Formulation}
\label{Sec:System Model and Problem Formulation}

\subsection{System Model}

We consider an edge video analytics system, where multiple devices collect video sensing data from the environment and forward them to an edge server to output the inference result.
Such edge video analytics can effectively overcome the limitations of resource-constrained devices, as the server has abundant computational resources to efficiently execute the inference task.
Besides, leveraging the spatial-temporal cues from multi-view videos overcomes the limited sensing capability of a single device, captures the dynamic patterns among consecutive frames, and improves inference performance.
Related application scenarios include multi-camera tracking \cite{ristani2016performance_duke_dataset} and vehicle re-identification \cite{liu2016deep_vehicle_reid}.
A sample system for pedestrian occupancy prediction \cite{hou2020multiview_multiview_detection} is shown in Fig. \ref{fig:sample_pedestrian_detection}.

The main challenge of edge inference systems is the limited communication bandwidth, which makes it infeasible to transmit high-volume raw videos to the edge server.
Besides, minimizing the communication overhead is also desirable to save device energy.
Thus, we will adopt the task-oriented communication principle \cite{letaief2021edge} that only transmits the information that is essential to the downstream tasks to the edge server for inference. 
Compared with image-based tasks investigated in \cite{shao2021learning,ShaoTWC}, video analytics tasks are more challenging since the input data are temporally correlated. 
The following subsections first present a probabilistic modeling of the edge video analytics system and then formulate the design problems for task-oriented communication.

\begin{table*}[t]
\small
\centering
\caption{Primary notations and descriptions.}
\begin{tabular}{l|l}
\hline
Notation & Description  \\ \hline
$\bm{x}_{t}^{(k)}$      &  Video frame of device $k$ at time step $t$. \\ \hline
$\bm{f}(\bm{x}_{t}^{(k)};\bm{\theta}^{(k)})$     & Feature extractor at device $k$ with parameters $\bm{\theta}^{(k)}$. \\ \hline
$\bm{z}_{t}^{(k)},\hat{\bm{z}}_{t}^{(k)}$ & Extracted feature and quantized feature. \\ \hline
$\bm{h}_{e}(\bm{z}_{t}^{(k)};\bm{\psi}_{e}^{(k)})$ & Hyper encoder at device $k$ with parameters $\bm{\psi}_{e}^{(k)}$. \\ \hline
$\bm{v}_{t}^{(k)},\hat{\bm{v}}_{t}^{(k)}$ & Hyper latent variable and quantized variable. \\ \hline
$q(\hat{\bm{v}}_{t}^{(k)};\bm{\psi}_{f}^{(k)})$ & Factorized entropy model parameterized by $\bm{\psi}_{f}^{(k)}$ \\ \hline
$\bm{h}_{d}(\bm{v}_{t}^{(k)};\bm{\psi}_{d}^{(k)})$ & Hyper decoder at device $k$ with parameters $\bm{\psi}_{d}^{(k)}$. \\ \hline
$q(\hat{\bm{z}}_{t}^{(k)}|\hat{\bm{v}}_{t}^{(k)};\bm{\psi}_{d}^{(k)})$ & Variational approximation to $p(\hat{\bm{z}}_{t}^{(k)}|\hat{\bm{v}}_{t}^{(k)})$.\\ \hline
$R_{\text{bit}}$ &  Bit constraint in loss $\mathcal{L}_{1}$.\\ \hline
$\tau_{1}$ &  Temporal parameter of the feature extractor and the temporal-spatial fusion module.\\ \hline
$\tau_{2}$ & Temporal parameter of the entropy model.\\ \hline
$\boldsymbol{g}(\hat{\boldsymbol{z}}_{t}^{(1: K)} ; \boldsymbol{\varphi}_{\tau})$ & The $\tau$-th auxiliary predictor with parameters $\boldsymbol{\varphi}_{\tau}$. \\ \hline
$q(\boldsymbol{y}_{t}, \ldots, \boldsymbol{y}_{t+\tau_{1}} | \hat{\boldsymbol{z}}_{t}^{(1: K)} ; \boldsymbol{\varphi}_{0: \tau_{1}})$ & 
Variational approximation to $p(\boldsymbol{y}_{t}, \ldots, \boldsymbol{y}_{t+\tau_{1}} | \hat{\boldsymbol{z}}_{t}^{(1: K)})$.
 \\ \hline
$\bm{h}_{p}(\hat{\boldsymbol{z}}_{t-1}^{(k)}, \ldots, \hat{\boldsymbol{z}}_{t-\tau_{1}}^{(k)} ; \boldsymbol{\psi}_{p}^{(k)})$ & Parametric transform with parameters $\boldsymbol{\psi}_{p}^{(k)}$ in the temporal entropy model. \\ \hline
$q(\hat{\boldsymbol{z}}_{t}^{(k)} | \hat{\boldsymbol{z}}_{t-1}^{(k)}, \ldots, \hat{\boldsymbol{z}}_{t-\tau_{2}}^{(k)} ; \boldsymbol{\psi}_{p}^{(k)})$ & Variational approximation to $p(\hat{\boldsymbol{z}}_{t}^{(k)} | \hat{\boldsymbol{z}}_{t-1}^{(k)}, \ldots, \hat{\boldsymbol{z}}_{t-\tau_{2}}^{(k)})$. \\ \hline
$\boldsymbol{g}(\hat{\boldsymbol{z}}_{t}^{(1: K)}, \ldots, \hat{\boldsymbol{z}}_{t-\tau_{1}}^{(1: K)} ; \boldsymbol{\phi})$ &  Spatial-temporal fusion module with parameters $\boldsymbol{\phi}$. \\ \hline
\end{tabular}
\end{table*}

\subsection{Probabilistic Modeling}
We consider an edge video analytics system comprising $K$ devices as shown in Fig. \ref{fig:edge_video_analytics_framework}.
The consecutive frames of each device $k \in 1:K$ with length $N$ are denoted as $\bm{x}_{1:N}^{(k)} := \{\bm{x}_{1}^{(k)},\ldots,\bm{x}_{N}^{(k)}\}$.
At each time step $t \in 1:N$, the input views captured by different devices are denoted by $\bm{x}_{t}^{(1:K)} = \{\bm{x}_{t}^{(1)},\ldots,\bm{x}_{t}^{(K)}\}$, which is deemed as realizations of random variables $X_{t}^{(1:K)} := \{{X}_{t}^{(1)},\ldots,{X}_{t}^{(K)}\}$.
The edge server aims to predict the desired output $\bm{y}_{1:N} := \{\bm{y}_{1},\ldots,\bm{y}_{N}\}$ based on the frames collected by the devices.
In particular, we define the random variables of the output as $Y_{1:N} := \{Y_{1},\ldots,Y_{N}\}$.
As not every part of the input data $\bm{x}_{t}^{(k)}$ is relevant to the inference task, the devices should identify the useful information $\bm{z}_{t}^{(k)}$ and discard the redundancy to reduce the communication overhead.
At each time step $t$, the device $k$ extracts the task-relevant feature $\bm{z}_{t}^{(k)}$ from its input data $\bm{x}_{t}^{(k)}$ and then rounds the feature to $\hat{\boldsymbol{z}}_{t}^{(k)}=\lfloor\boldsymbol{z}_{t}^{(k)}\rceil$.
Particularly, the extracted features $\bm{z}_{1:N}^{(k)} : = \{\bm{z}_{1}^{(k)},\ldots,\bm{z}_{N}^{(k)}\}$ and the discrete features $\bm{\hat{z}}_{1:N}^{(k)} := \{\bm{\hat{z}}_{1}^{(k)},\ldots,\bm{\hat{z}}_{N}^{(k)}\}$ are regarded as the instantiations of random variables $Z_{1:N}^{(k)} : = \{Z_{1}^{(k)},\ldots,Z_{N}^{(k)}\}$ and $\hat{Z}_{1:N}^{(k)} : = \{\hat{Z}_{1}^{(k)},\ldots\hat{Z}_{N}^{(k)}\}$, respectively.
At the network edge, the server receives the features from the devices to predict the target variables.
The key design objective is to minimize the communication overhead in feature transmission while providing satisfactory inference performance.

\subsection{Problem Formulation}

Learning a task-oriented communication framework for edge video analytics can be decomposed into three subproblems, including feature extraction, feature encoding, and joint inference.

\subsubsection{Feature Extraction}
At each time slot $t$, the extracted feature $\hat{\bm{z}}_{t}^{(k)}$ at device $k$ is transmitted to the edge server.
By extracting more task-relevant information from the input frame, the edge server holds the potential to improve the inference performance.
However, this comes at the cost of increased communication overhead.
So the extracted features should be maximally informative about the target variable $\bm{y}_{t}$ (i.e., to promote inference performance), subject to a constraint on its representation complexity (i.e., to control the communication overhead).
In this work, we resort to the deterministic information bottleneck to characterize the inherent tradeoff between the rate and inference performance.
Particularly, the mutual information $I(Y_{t};\hat{Z}_{t}^{(k)})$ is adopted to measure the informativeness between the target $Y_{t}$ and the feature $\hat{Z}_{t}^{(k)}$.
The entropy $H(\hat{Z}_{t}^{(k)})$ is utilized to represent the complexity of $\hat{Z}_{t}^{(k)}$.
In edge video analytics, the ideal feature $\hat{\bm{z}}_{t}^{(k)}$ should satisfy $I(Y_{t};\hat{Z}_{t}^{(k)}) = I(Y_{t};{X}_{\leq t}^{(k)})$, where ${X}_{\leq t}^{(k)} := \{{X}_{1}^{(k)},\ldots,{X}_{t}^{(k)}\}$.
This implies that the feature extractor needs to identify the task-relevant information from both the current frame as well as the previous frames.

However, given the limited resources at the devices, taking all the frames $\bm{x}_{\leq t}^{(k)}$ as inputs for feature extraction is computationally prohibitive and will also induce excessive memory consumption.
To alleviate this problem, we simplify the feature extraction process.
Each device $k \in 1:K$ extracts the feature $\hat{\bm{z}}_{t}^{(k)}$ only from the current frame $\bm{x}_{t}^{(k)}$ using the function $\hat{\bm{z}}_{t}^{(k)} = \lfloor\bm{f}(\bm{x}_{t}^{(k)};\bm{\theta}^{(k)})\rceil$ with parameters $\bm{\theta}^{(k)}$.
The discrete feature $\bm{\hat{z}}_{t}^{(k)}$ is expected to discard the task-irrelevant redundancy and preserve the information about the current target variable $\bm{y}_{t}$ as well as the temporal clues that help to predict the next $\tau_{1}$ variables $\{\bm{y}_{t+1},\ldots,\bm{y}_{t+\tau_{1}}\}$.
This suggests the objective:
\begin{equation}
\label{equ:multi_device_extraction}
    \min_{\bm{\theta}^{(1:K)}}  - \sum_{t=1}^{N} I(Y_{t},\ldots, Y_{t+\tau_{1}} ; \hat{Z}_{t}^{(1: K)}) + \beta \sum_{t=1}^{N} \sum_{k=1}^{K} H(\hat{Z}_{t}^{(k)}), 
\end{equation}
where the parameter $\beta >0$ trades off between the inference performance and the entropies of features, and $\bm{\theta}^{(1:K)}$ denotes the parameters $\{\bm{\theta}^{(1)},\ldots,\bm{\theta}^{(K)}\}$.
The first term represents the inference performance, and the second term denotes the amount of information preserved by the features.
As minimizing the objective function in (\ref{equ:multi_device_extraction}) requires high computation cost to estimate the distributions $p(\bm{y}_{t}|\bm{\hat{z}}_{t}^{(1:K)})$ and $p(\bm{\hat{z}}_{t}^{(k)})$, we adopt variational approximation \cite{balle2018variational1} to reformulate it into an amenable form in Section \ref{subsec:method_feature_extraction}.

\subsubsection{Feature Encoding}
After the feature extraction process, the second step is to encode the resulting discrete features for efficient transmission.
To losslessly reconstruct the features at the server, we adopt the entropy coding that relies on a probability model (i.e., an entropy model) of the discrete variable to estimate the marginal distribution of the features.
Denote the probability distribution of $\hat{\bm{z}}_{t}^{(k)}$ as $p(\hat{\bm{z}}_{t}^{(k)})$ and the estimated distribution of the entropy model as $q(\hat{\bm{z}}_{t}^{(k)})$.
Entropy coding is able to encode $\hat{\bm{z}}_{t}^{(k)}$ at the bitrate of the cross-entropy $H(p(\hat{\bm{z}}_{t}^{(k)}),q(\hat{\bm{z}}_{t}^{(k)}))$ with negligible overhead.
The better the entropy model predicts the distribution, the fewer bits are required to represent the discrete feature.
To effectively reduce the communication overhead in feature transmission, an intuitive method is to encode every subsequent feature $\hat{\bm{z}}_{t}^{(k)}$ depending on the previous features $\hat{\bm{z}}_{<t}^{(k)}$ to exploit the temporal dependence.
Therefore, we develop a temporal entropy model to estimate the conditional distribution of the current features given the previous ones.
To reduce the modeling complexity, we make a $\tau_{2}$-th order Markov assumption such that each feature $\hat{\bm{z}}_{t}^{(k)}$ for $k \in 1:K$ only depends on the previous $\tau_{2}$ features $\{\hat{\bm{z}}_{t-1}^{(k)},\ldots,\hat{\bm{z}}_{t-\tau_{2}}^{(k)}\}$, i.e., $p(\hat{\bm{z}}_{t}^{(k)}|\hat{\bm{z}}_{<t}^{(k)}) = p(\hat{\bm{z}}_{t}^{(k)}|\hat{\bm{z}}_{t-1}^{(k)},\ldots,\hat{\bm{z}}_{t-\tau_{2}}^{(k)})$.
Define $p(\hat{\bm{z}}_{t}^{(k)}|\hat{\bm{z}}_{t-1}^{(k)},\ldots,\hat{\bm{z}}_{t-\tau_{2}}^{(k)})$ and $q(\hat{\bm{z}}_{t}^{(k)}|\hat{\bm{z}}_{t-1}^{(k)},\ldots,\hat{\bm{z}}_{t-\tau_{2}}^{(k)};\bm{\psi}_{p}^{(k)})$ as the true and estimated temporal conditional distributions of $\bm{z}_{t}^{(k)}$.
In particular, the temporal entropy model of device $k$ is parameterized by $\bm{\psi}_{p}^{(k)}$.
The conditional cross-entropy 
$$H(p(\hat{\boldsymbol{z}}_{t}^{(k)} | \hat{\boldsymbol{z}}_{t-1}^{(k)}, \ldots, \hat{\boldsymbol{z}}_{t-\tau_{2}}^{(k)}), q(\hat{\boldsymbol{z}}_{t}^{(k)} | \hat{\boldsymbol{z}}_{t-1}^{(k)}, \ldots, \hat{\boldsymbol{z}}_{t-\tau_{2}}^{(k)} ; \boldsymbol{\psi}_{p}^{(k)}))$$ 
is expected to achieve a lower bitrate compared with the independent coding schemes.
In our framework, each device $k$ minimizes the conditional cross-entropy over the parameters $\bm{\psi}_{p}^{(k)}$ to reduce the communication overhead. More details about the training process have been deferred to Section \ref{sec:method_temporal_entropy_model}.

\subsubsection{Joint Inference}
\label{subsubsec:problem_formulation_fusion_model}

Once the edge server has received the extracted features from devices, it can output the inference result based on the conditional distribution of the target variable given these features.
As they are temporarily correlated, it is expected that jointly leveraging both the current received features $\hat{\bm{z}}_{t}^{(1:K)}$ as well as $\tau_{1}$ previous features $\{\hat{\bm{z}}_{t-\tau_{1}}^{(1:K)},\ldots,\hat{\bm{z}}_{t-1}^{(1:K)}\}$ reduces the uncertainty in inference.
However, the distribution $p(\bm{y}_{t}|\hat{\bm{z}}_{t-\tau_{1}}^{(1:K)},\ldots,\hat{\bm{z}}_{t}^{(1:K)})$ is difficult to compute due to the high-dimensional integral.
We apply approximate inference methods by setting up a spatial-temporal fusion module $\bm{\hat{y}}_{t} = \bm{g}(\hat{\bm{z}}_{t}^{(1:K)},\ldots,\hat{\bm{z}}_{t-\tau_{1}}^{(1:K)};\bm{\phi})$ with parameters $\bm{\phi}$, which takes the features $\{\hat{\bm{z}}_{t-\tau_{1}}^{(1:K)},\ldots,\hat{\bm{z}}_{t}^{(1:K)}\}$ as input to output the inference result $\hat{\bm{y}}_{t}$.
Let $d(\bm{y}_{t},\hat{\bm{y}}_{t};\bm{\phi})$ represent a metric to measure the distortion between the predicted output $\hat{\bm{y}}_{t}$ and the target $\bm{y}_{t}$.
The parameters $\bm{\phi}$ are optimized by minimizing this distortion function.
The detailed training process is illustrated in Section \ref{sec:method_temporal_fusion_model}.

\section{Task-Orientend Communication with Temporal Entropy Coding}

\label{Sec:Task-Orientend Communication with Temporal Entropy Coding}

We propose a task-oriented communication framework with a temporal entropy model (TOCOM-TEM) for edge video analytics.
It contains three components including task-relevant feature extraction, temporal entropy model, and spatial-temporal fusion module.
First, each device $k\in 1\!:\!K$ extracts the task-relevant feature $\bm{z}_{t}^{(k)}$ from the current input $\bm{x}_{t}^{(k)}$, and then applies entropy coding to compress the quantized feature $\hat{\bm{z}}_{t}^{(k)}$ to a bitstream before transmitting it to the edge server.
As the entropy coding relies on a probability model of the discrete variable, our method constructs a temporal entropy model that approximates the conditional distribution $p(\bm{\hat{z}}_{t}^{(k)}|\bm{\hat{z}}_{t-1}^{(k)},\ldots,\bm{\hat{z}}_{t-\tau}^{(k)})$ to reduce the bitrate by exploiting the relationships among features $\{\bm{\hat{z}}_{t}^{(k)},\ldots,\bm{\hat{z}}_{t-\tau_{2}}^{(k)}\}$.
Finally, a spatial-temporal fusion module at the edge server jointly leverages the current received features $\hat{\bm{z}}_{t}^{(k)}$ as well as the previous features $\{ \hat{\bm{z}}_{t-1}^{(1:K)},\ldots, \hat{\bm{z}}_{t-\tau_{1}}^{(k)} \}$ to predict the target variable $\hat{\bm{y}}_{t}$.
This section elaborates on the details of these three components.

\begin{figure*}[t]
\centering
\begin{minipage}[t]{1\textwidth}
\centering
\includegraphics[width=0.95\linewidth]{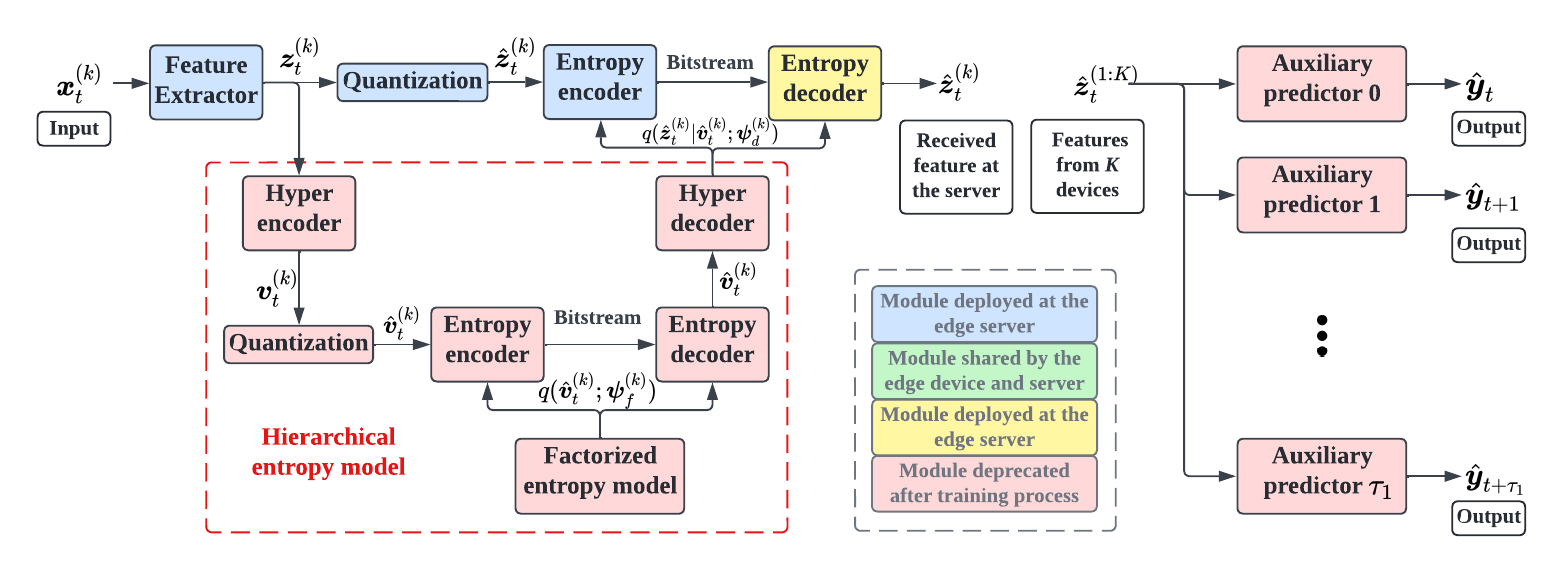}
\caption{\textbf{Task-relevant feature extraction.} Our method trains the feature extractors with hierarchical entropy models and auxiliary predictors.}
\label{Fig:phase1_feature_extraction}
\end{minipage}
\begin{minipage}[t]{1\textwidth}
\centering
\includegraphics[width=1\linewidth]{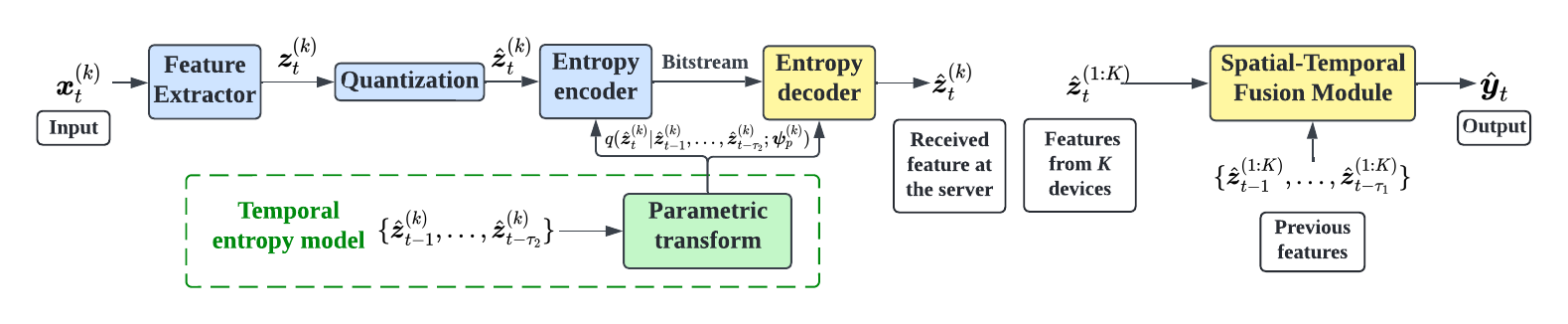}
\caption{\textbf{Temporal entropy model (left) and spatial-temporal fusion module (right).} 
Once the feature extractors have been optimized, the proposed method carries out two steps: (1) It trains temporal entropy models to reduce the communication overhead by exploiting the redundancy among consecutive features.
(2) It constructs a spatial-temporal fusion module to enhance inference performance by jointly utilizing the current and previous features.
}
\label{Fig:phase2_temporal_entropy_model_and_fusion_module}
\end{minipage}
\end{figure*}

\subsection{Task-Relevant Feature Extraction}
\label{subsec:method_feature_extraction}
The first step is to extract the task-relevant features from the video frames.
The loss function in (\ref{equ:multi_device_extraction}) aims at maximizing the informativeness of the extracted features, while minimizing its communication overhead.
However, directly optimizing the feature extractors based on this loss faces two challenges.
One is that the gradients with respect to the parameters $\bm{\theta}^{(1:K)}$ are zero almost everywhere due to the quantization.
The other is that taking the derivatives of the mutual information and the entropy is generally intractable since estimating the distributions $p(\bm{y}_{t},\ldots,\bm{y}_{t+\tau_{1}}|\hat{\bm{z}}_{t}^{(1:K)})$ and $p(\hat{\bm{z}}_{t}^{(k)})$ are computationally prohibitive.
In the following, we adopt \emph{continuous relaxation} and \emph{variational approximation} to alleviate these two problems, respectively.

\subsubsection{Continuous relaxation}
To allow optimization via gradient descent type algorithms in the presence of quantization, we follow the noise-based relaxation in \cite{balle2017end-to-end} to enable back-propagation.
The main idea is replacing the quantization operation with additive uniform noise $\mathcal{U}(-0.5, 0.5)$. 
Denote the $i$-th element of the extracted feature $\bm{z}_{t}^{(k)}$ as $z_{t,i}^{(k)}$.
The probability density function of $ \tilde{z}_{t,i}^{(k)} = z_{t,i}^{(k)} + \mathcal{U}(-0.5, 0.5)$ is a continuous relaxation of the probability mass function of $\hat{z}_{t,i}^{(k)} = \lfloor z_{t,i}^{(k)} \rceil$ such that $p({\tilde{z}}_{t,i}^{(k)} = \hat{z}) = p({\hat{z}}_{t,i}^{(k)} = \hat{z})$ for integer $\hat{z} \in \mathbb{Z}$.
This implies that the probability density function of $\tilde{\bm{z}}_{t}^{(k)}$ can be used to accurately calculate the entropy of $\hat{\bm{z}}_{t}^{(k)}$, and the differential entropy of variable $\bm{z}$ is a continuous proxy for the entropy of $\hat{\bm{z}}_{t}^{(k)}$ in the training process.

\subsubsection{Variational approximation}
Given the parameters $\bm{\theta}^{(1:K)}$, the distributions $p(\bm{y}_{t},\ldots,\bm{y}_{t+\tau_{1}}|\hat{\bm{z}}_{t}^{(1:K)})$ and $p(\hat{\boldsymbol{z}}_{t}^{(k)})$ are fully determined by the data distribution $p(\bm{x}_{1:N}^{(1:K)},\bm{y}_{1:N})$ and the feature extraction function $\bm{f}(\bm{x}_{t}^{(k)};\bm{\theta}^{(k)})$.
In cases where the data distribution is neither known perfectly nor can be estimated to high accuracy, minimizing the objective function in (\ref{equ:multi_device_extraction}) is nearly intractable.
In this situation, we resort to the variational approximation method to formulate a tractable lower bound.
The key idea is to posit a family of distributions and find a member of that family which is close to the target distribution.

To approximate the distribution $p(\bm{y}_{t},\ldots,\bm{y}_{t+\tau_{1}}|\hat{\bm{z}}_{t}^{(1:K)})$, we construct $\tau_{1} + 1$ auxiliary predictors $\bm{g}(\hat{\bm{z}}_{t}^{(1:K)};\bm{\varphi}_{\tau})$ for $\tau \in \{0,\ldots,\tau_{1}\}$.
As shown in Fig. \ref{Fig:phase1_feature_extraction}, these predictors infer the future output $\hat{\bm{y}}_{t+\tau}$ based on the features $\hat{\bm{z}}_{t}^{(1:K)}$.
The parameters $\bm{\varphi}_{0:\tau_{1}} := \{\bm{\varphi}_{0},\ldots,\bm{\varphi}_{\tau_{1}} \}$ represent the neural network layers.
Define loss function $d(\bm{y}_{t+\tau},\hat{\bm{y}}_{t+\tau};\bm{\varphi}_{\tau})$ that measures the discrepancy between the ground truth $\bm{y}_{t+\tau}$ and $\hat{\bm{y}}_{t+\tau}$.
We formulate the variational distribution as $q(\bm{y}_{t},\ldots,\bm{y}_{t+\tau_{1}}|\hat{\bm{z}}_{t}^{(1:K)};\bm{\varphi}_{0:\tau_{1}}) \propto \exp(- \sum_{\tau=0}^{\tau_{1}} w_{\tau}d(\bm{y}_{t+\tau},\hat{\bm{y}}_{t+\tau};\bm{\varphi}_{\tau}))$, where the scalars $\{w_{0},\ldots,w_{\tau_{1}}\}$ are weighting coefficients.
Note that the joint entropy $H(Y_{t},\ldots,Y_{t+\tau_{1}})$ is a constant.
Minimizing the cross entropy between the true and variational distributions is equivalent to minimizing the upper bound of $-I(Y_{t},\ldots, Y_{t+\tau_{1}} ; \hat{Z}_{t}^{(1: K)})$ in (\ref{equ:multi_device_extraction}).

Besides, we construct $K$ entropy models as variational approximations of the distributions $p(\hat{\bm{z}}_{t}^{(k)})$ for $k \in 1\!:\!K$.
These models are built over the hierarchical entropy model \cite{balle2017end-to-end}, which introduces a hyperprior to better capture the correlation among the elements in the extracted feature.
As shown in Fig. \ref{Fig:phase1_feature_extraction}, hyper encoder $\bm{h}_{e}^{(k)}({\bm{z}}_{t}^{(k)};\bm{\psi}_{e}^{(k)})$ transforms the extracted feature ${\bm{z}}_{t}^{(k)}$ to a hyper-latent $\bm{v}_{t}^{(k)}$, where $\bm{\psi}_{e}^{(k)}$ represents the neural network parameters.
Then, the device adopts entropy coding to encode the quantized hyper latent $\hat{\bm{v}}_{t}^{(k)}$ and transmits the compressed bitstream to the server.
Following \cite{balle2018variational1, minnen2018jointballe}, we adopt a fully factorized entropy model $q(\hat{\bm{v}}_{t}^{(k)};\bm{\psi}_{f}^{(k)})$ to approximate the marginal distribution $p(\hat{\bm{v}}_{t}^{(k)})$, where the parameters $\bm{\psi}_{f}^{(k)}$ correspond to the univariate non-parametric density model \cite{balle2018variational1}.
From the perspective of data compression, the hyperprior is the side information that allows the server to compress the feature $\hat{\bm{z}}_{t}^{(k)}$ based on a conditional entropy model $q(\hat{\bm{z}}_{t}^{(k)}|\hat{\bm{v}}_{t}^{(k)};\bm{\psi}_{d}^{(k)})$.
To ensure a good match between the quantized feature $\hat{\bm{z}}_{t}^{(k)}$ and the continuous-valued feature $\tilde{\bm{z}}_{t}^{(k)}$, we model each element $\hat{{z}}_{t,i}^{(k)}$ in feature $\hat{\bm{z}}_{t}^{(k)}$ as a Gaussian convolved with a unit uniform distribution $\mathcal{U}(-0.5, 0.5)$ as follows:
\begin{align*}
q({\hat{z}}_{t,i}^{(k)}|\hat{\bm{v}}_{t}^{(k)}) &  := \left( \mathcal{N}({\mu}_{i}^{(k)},{\sigma}_{i}^{(k)}) * \mathcal{U}(-0.5,0.5) \right) ({\hat{z}}_{t,i}^{(k)}),   \\
& \text{with } \bm{\mu}^{(k)},\bm{\sigma}^{(k)} = \bm{h}_{d}(\hat{\bm{v}}_{t}^{(k)};\bm{\psi}_{d}^{(k)}),
\end{align*}
where the parameters $\mu_{i}^{(k)}$ and $\sigma_{i}^{(k)}$ correspond to the $i$-th element in $\bm{\mu}^{(k)}$ and $\bm{\sigma}^{(k)}$, respectively.
The function $ \bm{h}_{d}(\hat{\bm{v}}_{t}^{(k)};\bm{\psi}_{d}^{(k)})$ is a hyper decoder parameterized by $\bm{\psi}_{d}^{(k)}$, which transforms the quantized latent $\hat{\bm{v}}_{t}^{(k)}$ to the mean and scale parameters $(\bm{\mu}^{(k)}, \bm{\sigma}^{(k)})$ of the Gaussian distribution. 
As both the quantized feature $\hat{\bm{z}}_{t}^{(k)}$ and the latent $\hat{\bm{v}}_{t}^{(k)}$ should be transmitted from the device to the server, the minimum communication cost is the joint entropy $H(\hat{Z}_{t}^{(k)},\hat{V}_{t}^{(k)})$.
With the conditional entropy model $q(\hat{\bm{z}}_{t}^{(k)}|\hat{\bm{v}}_{t}^{(k)};\bm{\psi}_{d}^{(k)})$ and the factorized entropy model $q(\hat{\bm{v}}_{t}^{(k)};\bm{\psi}_{f}^{(k)})$, we formulate a tractable upper bound of the joint entropy $H(\hat{\bm{z}}_{t}^{(k)},\hat{\bm{v}}_{t}^{(k)})$ as follows:
\begin{align}
    & H(\hat{Z}_{t}^{(k)},\hat{V}_{t}^{(k)}) =  H(\hat{Z}_{t}^{(k)}|\hat{V}_{t}^{(k)}) + H(\hat{V}_{t}^{(k)}), \notag \\
    \leq &  \mathbb{E} [ - \log q(\hat{\bm{z}}_{t}^{(k)}|\hat{\bm{v}}_{t}^{(k)};\bm{\psi}_{d}^{(k)})] + \mathbb{E} [ - \log q(\hat{\bm{v}}_{t}^{(k)};\bm{\psi}_{f}^{(k)})]. \label{equ:conditional_cross_entropy}
\end{align}
The first term on the right-hand side of the inequality represents the cost of transmitting $\hat{\bm{z}}_{t}^{(k)}$ by utilizing $\hat{\bm{v}}_{t}^{(k)}$ as the side information.
The second term corresponds to the overhead for transmitting the hyper latent $\hat{\bm{v}}_{t}^{(k)}$.
To control the communication cost in edge inference, we introduce an extra parameter $R_{\text{bit}}$ representing the bit constraint.
Our objective is to penalize the communication cost when it is larger than the given $R_{\text{bit}}$.
With the above preparation, we formulate a tractable loss function $\mathcal{L}_{1}$ in (\ref{equ:loss_feature_extraction}).
Minimizing it is equivalent to minimizing an upper bound of (\ref{equ:multi_device_extraction}).
The details of derivations are deferred to Appendix \ref{appendix:loss_1}.

\begin{figure*}[t]
\normalsize
\begin{align}
& \mathcal{L}_{1} := \sum_{t=1}^{N} \mathbb{E} \Biggl\{  - \log q(\bm{y}_{t},\ldots,\bm{y}_{t+\tau_{1}}|\hat{\bm{z}}_{t}^{(1:K)};\bm{\varphi}_{0:\tau_{1}})   + \beta \cdot \max \left\{ \sum_{k=1}^{K} \left[ - \log q(\hat{\bm{z}}_{t}^{(k)}|\hat{\bm{v}}_{t}^{(k)};\bm{\psi}_{d}^{(k)}) - \log q(\hat{\bm{v}}_{t}^{(k)};\boldsymbol{\psi}_{f}^{(k)}) \right], R_{\text{bit}}  \right\} \Biggl\}, \label{equ:loss_feature_extraction} \\
& \mathcal{L}_{2}^{(k)} := \sum_{t=1}^{N} H(p(\hat{\boldsymbol{z}}_{t}^{(k)} | \hat{\boldsymbol{z}}_{t-1}^{(k)}, \ldots, \hat{\boldsymbol{z}}_{t-\tau_{2}}^{(k)}), q(\hat{\boldsymbol{z}}_{t}^{(k)} | \hat{\boldsymbol{z}}_{t-1}^{(k)}, \ldots, \hat{\boldsymbol{z}}_{t-\tau_{2}}^{(k)} ; \boldsymbol{\psi}_{p}^{(k)})), \quad \quad  \quad \quad   \mathcal{L}_{3} := \mathbb{E}\left[ \sum_{t=1}^{N} d\left(\boldsymbol{y}_{t}, \hat{\boldsymbol{y}}_{t};\bm{\phi}\right)\right].
\label{equ:loss2and3}
\end{align}
\end{figure*}

\begin{algorithm*}[t]
\small
\caption{Training procedures of the proposed method}
\begin{algorithmic}[1]
\Require Training dataset, initialized parameters $\bm{\theta}^{(k)}$, $\bm{\psi}_{e}^{(k)}$, $\bm{\psi}_{d}^{(k)}$, $\bm{\psi}_{f}^{(k)}$, $\bm{\psi}_{p}^{(k)}$ for $k \in 1\!:\!K$, and $\bm{\varphi}_{0:\tau_{1}}$, $\bm{\phi}$.
\Ensure The optimized parameters $\bm{\theta}^{(k)}$, $\bm{\psi}_{p}^{(k)}$ for $k \in 1\!:\!K$, and $\bm{\phi}$.
\Repeat \ {\color{gray} \# Train the feature extractors along with hierarchical entropy models and auxiliary predictors.}
\State Randomly select a minibatch from the training dataset.
\While{$k=1$ to $K$}
\State Extract the features by the feature extractor of device $k$ with parameter $\bm{\theta}^{(k)}$. 
\State Compress the features based on the hierarchical entropy model with parameters $\{\bm{\psi}_{e}^{(k)},\bm{\psi}_{d}^{(k)},\bm{\psi}_{f}^{(k)} \}$.
\EndWhile
\State Output the inference results by the auxiliary predictors with parameters $\bm{\varphi}_{0:\tau_{1}}$.
\State Compute the empirical estimation of the loss function $\mathcal{L}_{1}$ in (\ref{equ:loss_feature_extraction}).
\State Update parameters $\bm{\theta}^{(k)}$, $\bm{\psi}_{e}^{(k)}$, $\bm{\psi}_{d}^{(k)}$, $\bm{\psi}_{f}^{(k)}$ for $k \in 1\!:\!K$, and $\bm{\varphi}_{0:\tau_{1}}$ through backpropagation.
\Until{Convergence of parameters $\bm{\theta}^{(k)}$, $\bm{\psi}_{e}^{(k)}$, $\bm{\psi}_{d}^{(k)}$, $\bm{\psi}_{f}^{(k)}$ for $k \in 1\!:\!K$ and $\bm{\varphi}_{0:\tau_{1}}$.}
\Repeat \ {\color{gray} \# Train the entropy temporal models and the spatial-temporal fusion model in parallel.}
\State Randomly select a minibatch from the training dataset.
\While{$k=1$ to $K$}
\State Extract the features by the feature extractor of device $k$ with parameter $\bm{\theta}^{(k)}$. 
\State Compress the features based on the temporal entropy model with parameters $\bm{\psi}_{p}^{(k)}$.
\State Compute the empirical estimation of the loss function $\mathcal{L}_{2}^{(k)}$ in (\ref{equ:loss2and3}).
\State Update parameters $\bm{\psi}_{p}^{(k)}$ through backpropagation.
\EndWhile
\State Output the inference results by the spatial-temporal fusion module with parameters $\bm{\phi}$.
\State Compute the empirical estimation of the loss function $\mathcal{L}_{3}$ in (\ref{equ:loss2and3}).
\State Update parameters $\bm{\phi}$ through backpropagation.
\Until{Convergence of parameters $\bm{\psi}_{p}^{(k)}$ for $k \in 1\!:\!K$ and $\bm{\phi}$.}
\end{algorithmic}
\label{algorithm:VIB}
\end{algorithm*}

\begin{remark}
The parameter $R_{\text{bit}}$ serves as a constraint on communication overhead.
The loss $\mathcal{L}_{1}$ will penalize the cost $\sum_{k=1}^{K} \left[ - \log q(\hat{\bm{z}}_{t}^{(k)}|\hat{\bm{v}}_{t}^{(k)};\bm{\psi}_{d}^{(k)}) - \log q(\hat{\bm{v}}_{t}^{(k)};\boldsymbol{\psi}_{f}^{(k)}) \right]$ only when it is larger than the given $R_{\text{bit}}$.
Otherwise, the second term of $\mathcal{L}_{1}$ will not affect the optimization.
Therefore, the training process is less sensitive to the choice of coefficient $\beta$.
In Section \ref{sec:method_temporal_entropy_model}, we will develop a temporal entropy model to further minimize the communication overhead by reducing the redundancy among consecutive features.
\end{remark}

\subsection{Temporal Entropy Model}
\label{sec:method_temporal_entropy_model}

Considering the temporal correlation in video frames, it is expected that encoding every extracted feature by using the previous ones as side information achieves a lower bitrate.
As shown in Fig. \ref{Fig:phase2_temporal_entropy_model_and_fusion_module}, we develop temporal entropy models to exploit the temporal redundancy.
Let $q({\hat{z}}_{t,i}^{(k)}|\bm{\hat{z}}_{t-1}^{(k)},\ldots,\bm{\hat{z}}_{t-\tau_{2}}^{(k)};\bm{\psi}_{p}^{(k)})$ be a variational approximation to the true conditional distribution.
We model it using a Gaussian distribution:

\begin{equation*}
\begin{aligned}
    & q({\hat{z}}_{t,i}^{(k)}|\bm{\hat{z}}_{t-1}^{(k)},\ldots,\bm{\hat{z}}_{t-\tau_{2}}^{(k)};\bm{\psi}_{p}^{(k)}) \\
    & = \left( \mathcal{N}({\mu}_{i}^{(k)},{\sigma}_{i}^{(k)}) * \mathcal{U}(-0.5,0.5) \right) ({\hat{z}}_{t,i}^{(k)}),
\end{aligned}
\end{equation*}
where we apply a parametric transform $\bm{\mu}^{(k)},\bm{\sigma}^{(k)} = \bm{h}_{p}(\bm{\hat{z}}_{t-1}^{(k)},\ldots,\bm{\hat{z}}_{t-\tau_{2}}^{(k)};\bm{\psi}_{p}^{(k)})$ to predict the mean and variance of the Gaussian distribution\footnote{With slight abuse of notations, we reuse $\bm{\mu}^{(k)},\bm{\sigma}^{(k)}$ as the output of the parametric transform.}.
The parameters $\bm{\psi}_{p}^{(k)}$ correspond to the weights of neural network layers.
Encoding the features based on the variational distribution results in a bitstream length that is roughly equal to the cross entropy between the true and variational distributions.
To reduce the communication overhead, we minimize the discrepancy between the variational and true distributions based on the cross-entropy loss $\mathcal{L}_{2}^{(k)}$ defined in (\ref{equ:loss2and3}).

\subsection{Spatial-Temporal Fusion Module}
\label{sec:method_temporal_fusion_model}

As the extracted features in our system try to capture the temporal clues that are relevant for the future target variables $\{\bm{y}_{t+1},\ldots,\bm{y}_{t+\tau_{1}}\}$, we proposed a spatial-temporal fusion module $\boldsymbol{g}(\hat{\boldsymbol{z}}_{t}^{(1: K)}, \ldots, \hat{\boldsymbol{z}}_{t-\tau_{1}}^{(1: K)} ; \boldsymbol{\phi})$.
As shown in Fig. \ref{Fig:phase2_temporal_entropy_model_and_fusion_module}, this module outputs the inference result $\hat{\boldsymbol{y}}_{t}$ by taking both the current received features $\hat{\boldsymbol{z}}_{t}^{(1: K)}$ as well as the previous $\tau_{1}$ features $\{\hat{\bm{z}}_{t-\tau_{1}}^{(1:K)},\ldots,\hat{\bm{z}}_{t-1}^{(1:K)}\}$ as input.
To optimize the neural network parameters $\boldsymbol{\phi}$, we introduce the loss function $\mathcal{L}_{3}$ in (\ref{equ:loss2and3}), which quantifies the distortion between the target variable $\bm{y}_{t}$ and the inference output $\hat{\bm{y}}_{t}$.

To train the feature extractors, temporal entropy models, and the spatial-temporal fusion module according to the loss functions in (\ref{equ:loss_feature_extraction}) and (\ref{equ:loss2and3}), we apply the Monte Carlo sampling method that randomly selects a mini-batch from the training dataset to obtain an unbiased estimate of the gradients.
The detailed training procedures are summarized in Algorithm 1.

\section{Performance Evaluation}
\label{Sec:Experiment}

In this section, we evaluate the performance of the task-oriented communication framework on two edge video analytics tasks, namely, multi-camera pedestrian occupancy prediction and multi-camera object detection\footnote{The code is available at \url{https://github.com/shaojiawei07/TOCOM-TEM}.}.

\subsection{Experimental Setup}

\subsubsection{Tasks and datasets}

We conduct the multi-camera pedestrian occupancy prediction on the Wildtrack dataset \cite{wildtrack_dataset}.
This dataset includes 400 synchronized frames from 7 cameras with overlapping field-of-view, covering a 12 meters by 36 meters ground plain.
The objective of this task is to estimate the probabilities of occupancy on the ground plane by exploiting the geometrical constraints from multiple views.
For annotation, the ground plane is quantized into a 480 
$\times$ 1440 grid, where each grid cell is a 2.5-centimeter square. 
Besides, we select the EPFL MVMC Detection dataset \cite{MVMC_dataset} to conduct the multi-camera object detection, which contains 6 sequences of frames including 1297 persons, 3553 cars, and 56 buses.
The task objective is to assign bounding boxes to describe the spatial location of semantic objects.
In the following experiments, we consider that each camera is deployed on an NVIDIA Jetson Orin NX board, and the edge server is equipped with an adequate number of NVIDIA A5000 Tensor Core GPUs.

\subsubsection{Metric}
We mainly investigate the rate-performance tradeoff.
The rate refers to the upload overhead, which is calculated by multiplying the average communication cost per frame by the number of cameras.
For the multi-camera pedestrian occupancy prediction task, we report the multiple object detection accuracy (MODA) as the performance indicator \cite{MODA_metric}.
This metric accounts for the normalized missed detections and false detections.
To evaluate the performance of multi-camera object detection, we adopt the COCO-style mean average precision (mAP)\footnote{We report the average mAP over different intersection over union (IoU) thresholds, which ranges from 0.5 to 0.95 with a step size of 0.05.} to measure the quality of the estimated bounding boxes \cite{lin2014microsoft_coco}.

\subsubsection{Proposed methods}
We evaluate the TOCOM-TEM method in the experiments.
Particularly, when the parameter $\tau_2$ is set to zero, we adopt the hierarchical entropy model for feature compression, i.e., encoding the current feature independently without leveraging other features as side information.
Besides, when setting the parameter $\tau_2 \geq 1$, we adopt the temporal entropy model to compress the extracted features.
In the multi-view pedestrian occupancy prediction task, the bit constraint $R_{\text{bit}}$ varies between $10^{5}$ and $5 \times 10^{6}$ to control the tradeoff between the communication cost and the performance.
For the multi-view object detection task, we set the bit constraint $R_{\text{bit}}$ in the range of $5 \times 10^{4}$ to $10^{6}$. 
The parameter $\beta$ in these two tasks is set to $10^{-5}$.

\subsubsection{Baselines} 
In this study, we compare the proposed methods against six different data-oriented communication methods.
The deep learning models are trained on the raw images and tested on the compressed data.
\begin{itemize}
\item Image compression: We select two conventional image compression methods, JPEG and WebP \cite{si2016research_webp}, and two deep image compression (DIC) methods with a factorized prior model \cite{balle2017end-to-end} and a scale hyperprior model \cite{balle2018variational1} as baselines.
The two learning-based methods use neural networks to learn an encoder-decoder pair for image compression and decompression.
The encoder outputs a quantized feature as a compact representation of the input image.
An entropy model is learned in an end-to-end manner to reduce the bitrate of the quantized feature by entropy coding.
In the experiments, the two DIC methods are implemented based on the CompressAI package \cite{begaint2020compressai_package}.
The compression ratios achieved by these methods are controlled by adjusting the value assigned to the corresponding \texttt{quality} parameter.

\item Video compression: As video frames usually contain abundant amounts of spatial and temporal redundancy, video compression algorithms attempt to store information more compactly.
We adopt two well-known video compression algorithms, Advanced Video Coding (AVC) \cite{richardson2011h_H264} and High Efficiency Video Coding (HEVC) \cite{sullivan2012overview_HECV}, as baselines for comparison.
The constant rate factor controls the compression level of video files. 
\end{itemize}
We evaluate these baseline methods in terms of the rate-performance curve by setting different quality values.

\begin{figure}[t]
\centering
\begin{minipage}[t]{1\linewidth}
\centering
\includegraphics[width=0.95\linewidth]{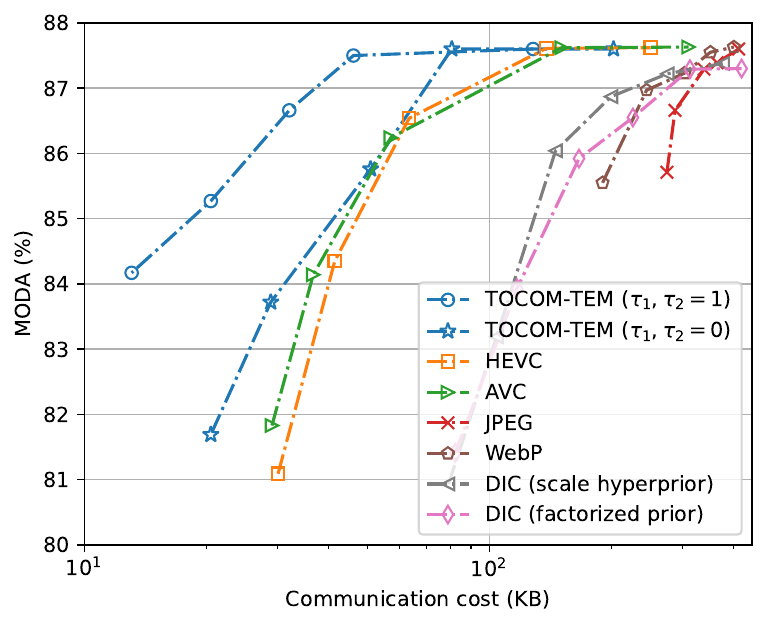}
\caption{The rate-performance curves of different methods in the multi-camera pedestrian occupancy prediction task.}
\label{fig:Final_wildtrack_MODA_bitrate_tradeoff}
\end{minipage}
\begin{minipage}[t]{1\linewidth}
\centering
\includegraphics[width=1\linewidth]{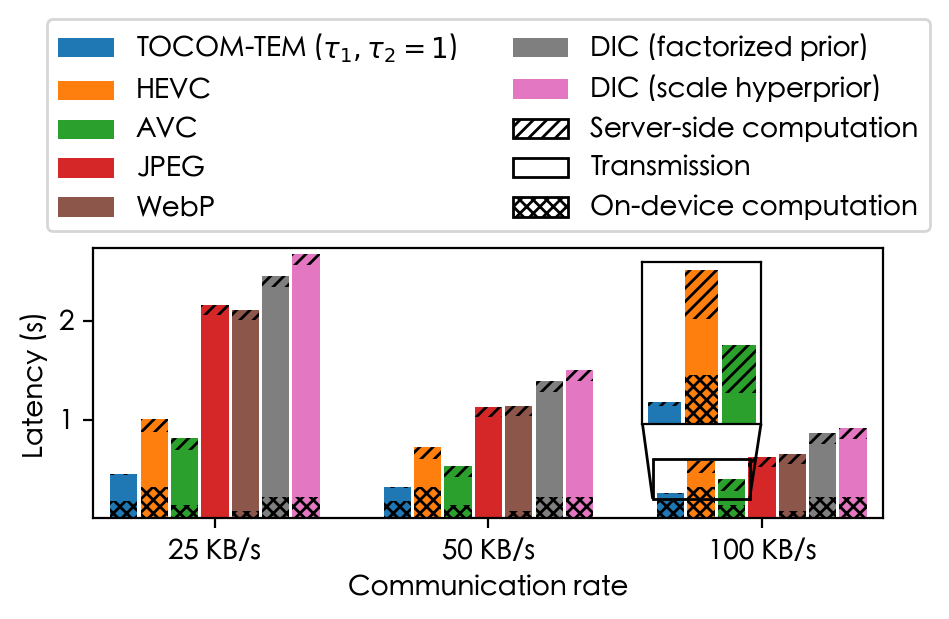}
\caption{Average inference latency per frame under different communication rates in the multi-camera pedestrian occupancy prediction task. The overall latency comprises on-device computation time, transmission latency, and server-side computation time. All the methods achieve a MODA of around 87\%.}
\label{fig:wildtrack_system_latency}
\end{minipage}
\end{figure}

\subsubsection{Neural network architecture}

In the multi-camera pedestrian occupancy prediction task, we adopt the MVDet model \cite{hou2020multiview_multiview_detection} as the backbone to estimate the pedestrian position by incorporating multiple camera views.
The MVDet model is composed of two main components: a feature extractor and a spatial aggregator. 
The feature extractor is similar to ResNet-18 \cite{ResNet}, but the last two layers are removed and the last three strided convolutions are replaced by dilated convolutions.
The spatial aggregator consists of three convolutional layers to exploit the spatial neighbor information of multiple views.
Besides, the hierarchical entropy model adopts five convolutional layers and five deconvolutional layers as a hyper encoder and a hyper decoder, respectively.
The temporal entropy model employs three convolutional layers as the parametric transform.

For the multi-camera detection task, we select the fully convolutional one-stage object detector (FCOS) \cite{tian2019fcos} to solve object detection in an anchor-free manner.
The FCOS framework utilizes a ResNet-50 model and a feature pyramid network (FPN) \cite{lin2017feature_FPN} for feature extraction.
An anchor-free detector outputs the bounding boxes based on the multi-level features.
Besides, we construct a residual building block consisting of three convolutional layers to fuse the feature maps from different cameras to exploit the spatial correlation.
To effectively compress the multi-scale features, the hierarchical entropy model contains four convolutional layers and four deconvolutional layers, and the temporal entropy model consists of five convolutional layers.

\begin{figure}[t]
\centering
\begin{minipage}[t]{1\linewidth}
\centering
\includegraphics[width=0.95\linewidth]{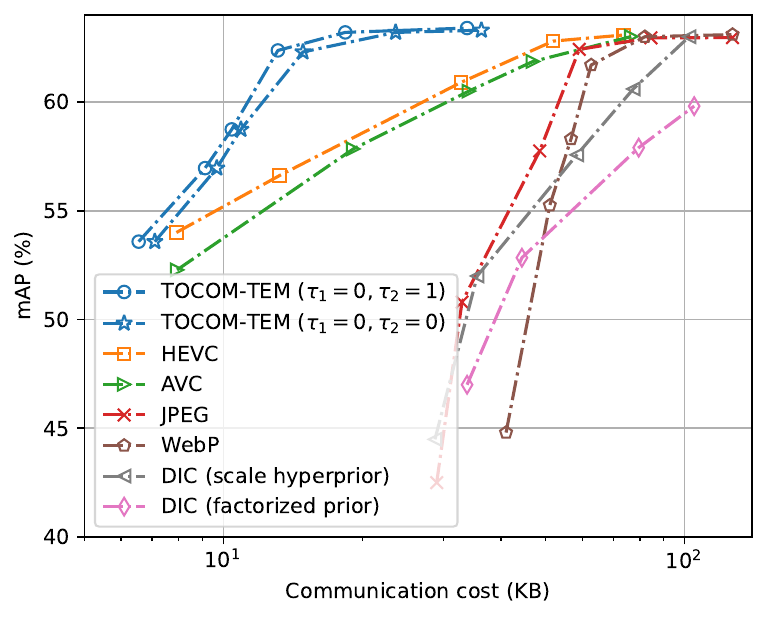}
\caption{The rate-performance curves of different methods in the multi-camera object detection task.
}
\label{fig:Final_EPFL_Accuracy_bitrate_tradeoff}
\end{minipage}
\begin{minipage}[t]{1\linewidth}
\centering
\includegraphics[width=1\linewidth]{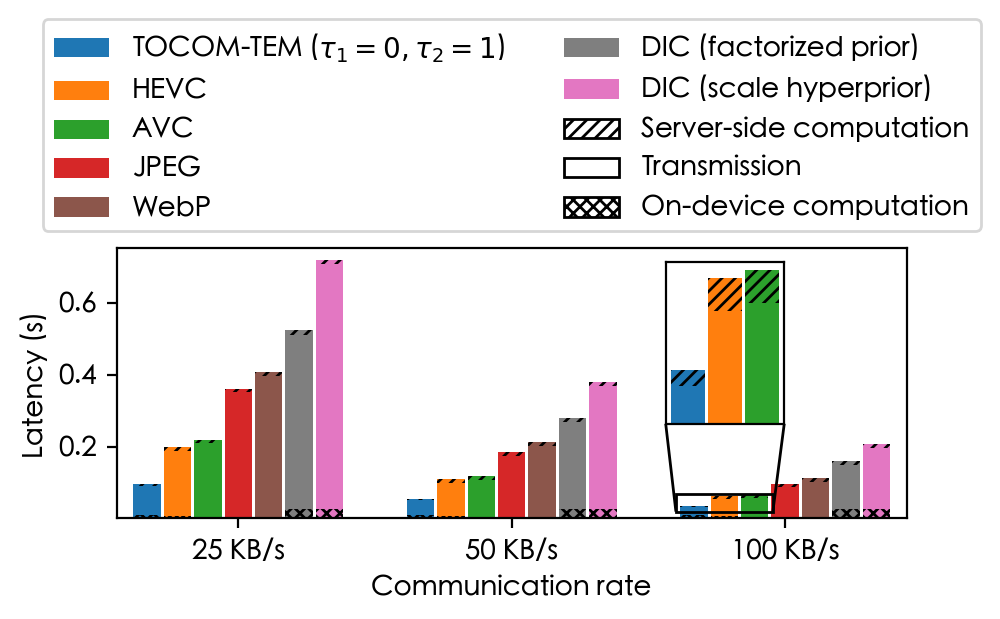}
\caption{Average inference latency per frame under different communication rates in the multi-camera object detection task. The overall latency comprises on-device computation time, transmission latency, and server-side computation time. All the methods achieve an mAP of around 60\%.}
\label{fig:epfl_system_latency}
\end{minipage}
\end{figure}

\subsection{Multi-Camera Pedestrian Occupancy Prediction}

This task aims at incorporating multiple camera views to predict pedestrian occupancy.
The multi-camera system alleviates the impact of occlusions in crowded scenes, but it induces much higher communication overhead compared with the single-camera system.
Fig. \ref{fig:Final_wildtrack_MODA_bitrate_tradeoff} presents the rate-performance curves of different methods for multi-camera pedestrian occupancy prediction, where the horizontal axis shows the communication cost, and the vertical axis corresponds to the inference performance.
For data-oriented communication, the result shows that video compression methods, i.e., AVC and HEVC, consistently outperform the image compression methods. This is because video compression methods can reduce the redundancy among consecutive frames to achieve a higher compression ratio.
Among all the configurations in the figure, our TOCOM-TEM method performs the best.
This demonstrates the effectiveness of the task-oriented communication framework that discards irrelevant information while maintaining inference performance.
Particularly, setting the parameters $\tau_{1},\tau_{2}$ to 1 in TOCOM-TEM achieves a better rate-performance tradeoff compared with setting them to 0.
This is attributed to the ability of the entropy model and spatial-temporal fusion module to exploit the temporal correlation among features.
When encoding feature $\hat{\bm{z}}_{t}^{(k)}$, the temporal entropy model adopts the previous feature $\hat{\bm{z}}_{t-1}^{(k)}$ as side information to reduce the bitrate.
Besides, the spatial-temporal fusion module reduces the uncertainty in prediction by jointly utilizing the current received features $\hat{\bm{z}}_{t}^{(1:K)}$ and the previous features $\hat{\bm{z}}_{t-1}^{(1:K)}$.

Next, we conduct a comparison of the latency across different methods, where the overall latency in edge inference comprises on-device computation time, transmission latency, and server-side computation time.
In order to prevent over-compression of input data, we adjust the compression ratio of each method to ensure that the MODA score is approximately 87\%.
Fig. \ref{fig:wildtrack_system_latency} depicts the inference latency at various communication rates.
In many cases, the transmission process is the main bottleneck causing delays in edge inference, especially when the communication rate is low.
The extra feature extraction step in our method increases the complexity on the device side, but it effectively removes the task-irrelevant information and largely reduces the communication overhead.
Therefore, the proposed TOCOM-TEM method consistently exhibits the smallest latency at various communication rates.

\begin{figure}[t]
\centering
\begin{minipage}[t]{1\linewidth}
\centering
\includegraphics[width=0.9\linewidth]{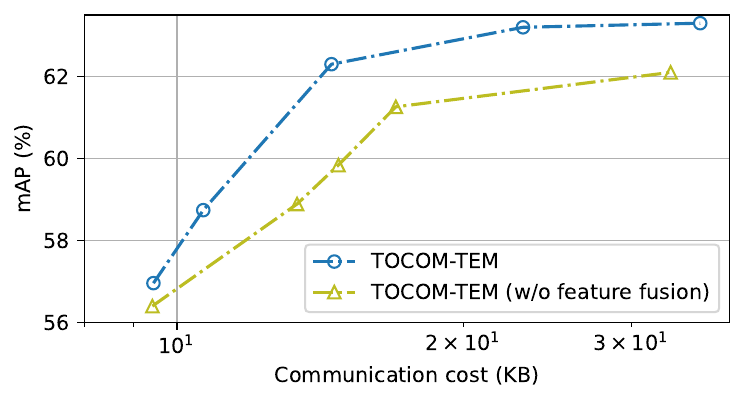}
\caption{The rate-performance curves of the proposed TOCOM-TEM method with and without the feature fusion module.}
\label{Fig:TOCOM-TEM_with_without_fusion}
\end{minipage}
\begin{minipage}[t]{1\linewidth}
\centering
\includegraphics[width=0.9\linewidth]{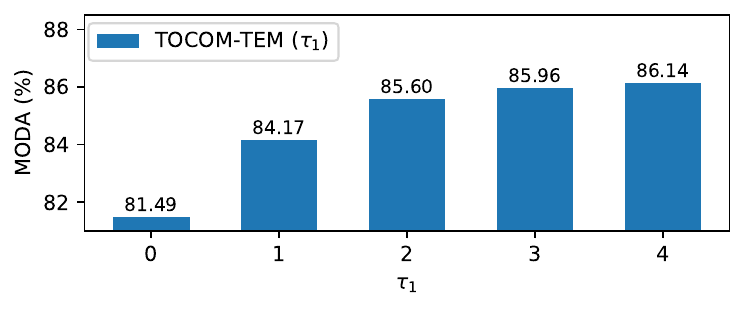}
\caption{Impact of the value of the parameter $\tau_{1}$ on the inference performance of the multi-camera pedestrian prediction task.}
\label{Fig:impact_tau_1}
\end{minipage}
\end{figure}

\begin{figure*}[t]
\centering
\subfigure[Multi-camera pedestrian occupancy prediction]{
\centering
\includegraphics[width=0.4\textwidth]{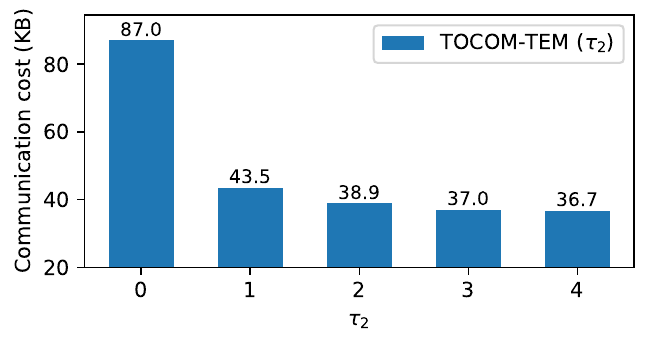}
}
\subfigure[Multi-camera object detection]{
\centering
\includegraphics[width=0.4\textwidth]{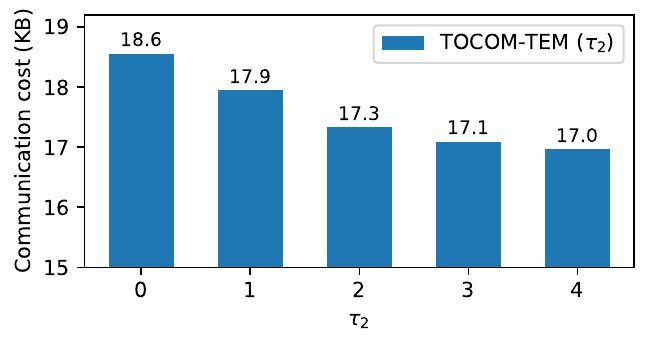}
}
\caption{Impact of the value of the parameter $\tau_{2}$ on the communication cost in (a) the multi-camera pedestrian occupancy prediction task and (b) the multi-camera object detection task.}
\label{Fig:impact_tau2_overview}

\end{figure*}

\begin{figure*}[t]
    \centering
    \includegraphics[width = 0.99\linewidth]{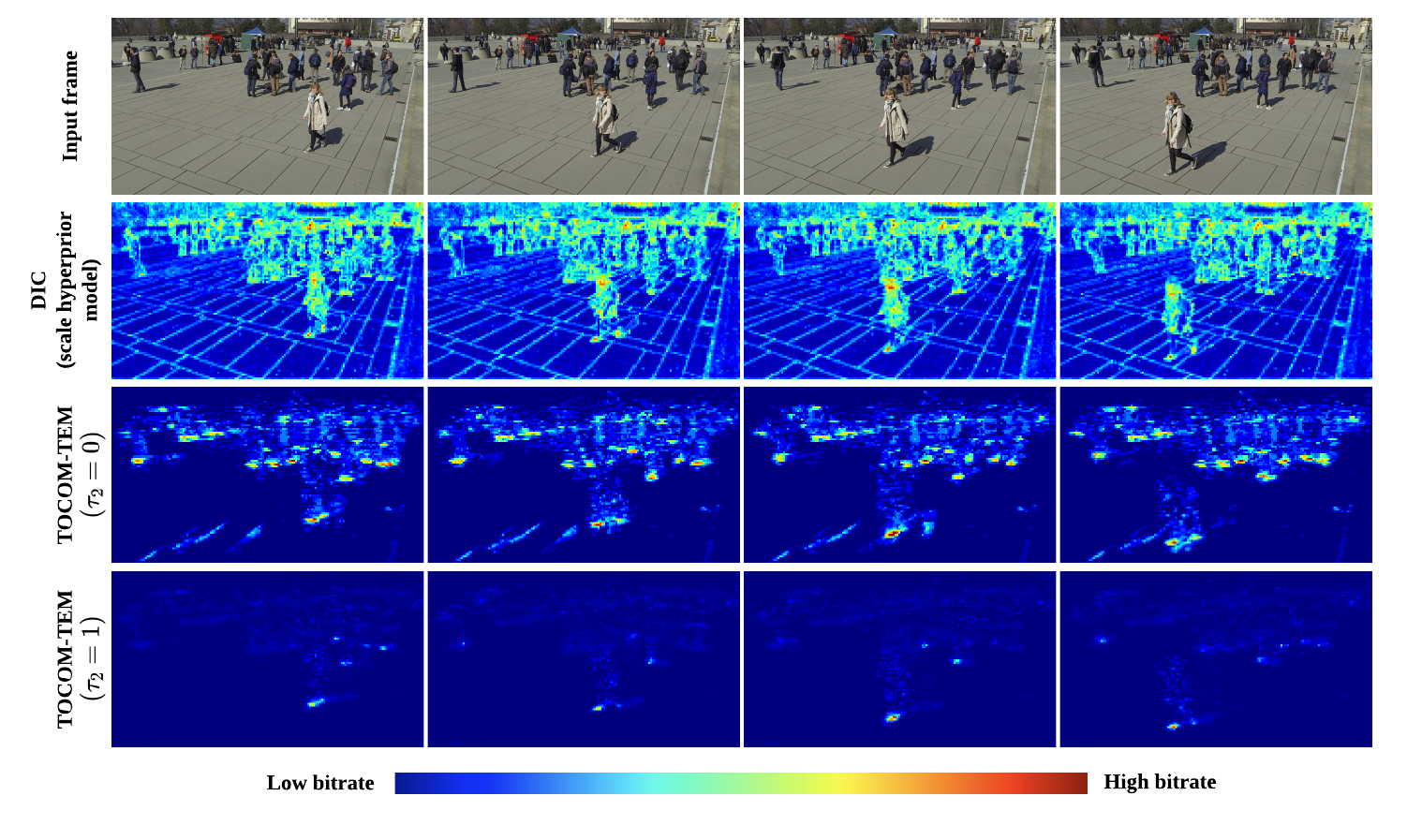}
    \caption{Bit allocation of the transmitted features of different methods for the multi-camera pedestrian occupancy prediction task.
    The data-oriented approach (the second row) allocates many bits to represent the background and ground texture to maintain the quality of the reconstructed images. 
    Our proposed TOCOM-TEM method (the third and fourth rows) mainly focuses on task-relevant information (e.g., the foot points of pedestrians) and discards the redundancy. 
    Specifically, using the temporal entropy model ($\tau_{2} = 1$) further reduces the communication overhead by exploiting the temporal correlation among frames.}
    \label{fig:WILDTRACK_entropy_model_visualization_example}
\end{figure*}

\subsection{Multi-Camera Object Detection}

The object detection task in a multi-camera environment enjoys the benefit of the images taken from different angles, which helps to take into account occlusions between objects for better bounding box regression.
We compare the rate-performance tradeoff of different methods, and the results are shown in Fig. \ref{fig:Final_EPFL_Accuracy_bitrate_tradeoff}.
The data-oriented communication methods (i.e., image compression and video compression) lead to higher communication cost compared with the task-oriented communication method.
This is attributed to the fact that data-oriented communication simply guarantees the correct reception of every single transmitted bit without optimizing the communication strategy for the downstream task, so a large amount of the transmitted data may be irrelevant to the inference task.
In our TOCOM-TEM method, using the temporal entropy model ($\tau_{2} = 1$) achieves a better rate-performance tradeoff compared with the hierarchical entropy model ($\tau_{2} = 0$).
This is because using the previous features as the side information in entropy coding further reduces the encoded data size.
Besides, Fig. \ref{fig:epfl_system_latency} displays the overall latency of various methods that achieve approximately 60\% mAP.
While our method introduces additional complexity on the device side due to the complex feature extraction process, the proposed TOCOM-TEM method still enables low-latency inference.
This is because the increase in computation latency is significantly outweighed by the reduction in transmission latency.

\subsection{Ablation Study}

In this subsection, we evaluate the impact of the feature fusion module and the parameters $\tau_{1},\tau_{2}$ of the proposed TOCOM-TEM method on the edge video analytics systems.
Besides, we discuss the bit allocation of the transmitted features in data-oriented communication and task-oriented communication.

\subsubsection{Impact of the feature fusion module}
To evaluate the effectiveness of the feature fusion module in our method, we consider a baseline denoted as (TOCOM-TEM w/o feature fusion), which does not exploit the correlation among multi-camera features in the inference and is in essence a method for single-camera inference.
We compare the rate-performance curves of the TOCOM-TEM method with and without the feature fusion module in the multi-camera object detection task.
In this experiment, the parameters $\tau_{1}$ and $\tau_{2
}$ are set to 0 and 1, respectively.
As shown in Fig. \ref{Fig:TOCOM-TEM_with_without_fusion}, the proposed TOCOM-TEM method achieves a better rate-performance tradeoff.
This is due to the feature fusion module that improves performance under occlusion by leveraging the spatial dependency among multiple cameras.

\subsubsection{Impact of the parameter $\tau_1$}

We investigate the performance of the TOCOM-TEM method with different values of $\tau_{1}$ in the multi-camera pedestrian occupancy prediction task.
The bit constraint $R_{\text{bit}}$ is set to $2.5 \times 10^{5}$ in this experiment.
As shown in Fig. \ref{Fig:impact_tau_1}, increasing the value of the parameter $\tau_{1}$ improves the MODA score.
This result is consistent with the analysis in Section \ref{subsubsec:problem_formulation_fusion_model} that leveraging both the current received features as well as $\tau_{1}$ previous features reduces the uncertainty in the prediction. 

\subsubsection{Impact of the parameter $\tau_2$}
To investigate the performance of the temporal entropy model in the proposed task-oriented communication framework, we evaluate the communication cost of our method by varying the value of parameter $\tau_{2}$.
Particularly, we set the bit constraint $R_{\text{bit}}$ to $10^{6}$ and $2 \times 10^{5}$ in the multi-camera pedestrian occupancy task and the multi-camera object detection task, respectively.
The temporal entropy model encodes the current feature $Z_{t}^{(k)}$ conditional on the previous $\tau_{2}$ features $\{Z_{t-1}^{(k)},\ldots, Z_{t-\tau_{2}}^{(k)}\}$.
As shown in Fig. \ref{Fig:impact_tau2_overview}, increasing the value of $\tau_{2}$ leads to a reduction in communication costs. 
These results align with the information theory that the conditional entropy $H(\hat{Z}_{t}^{(k)}|\hat{Z}_{t-1}^{(k)}, \ldots, \hat{Z}_{t-\tau_{2}}^{(k)})$ decreases monotonically as $\tau_{2}$ increases.

\subsubsection{Bit allocation in feature transmission}
This part discusses the difference between the extracted features in the DIC (scale hyperprior model) method and our TOCOM-TEM method.
To achieve the same MODA of around 85\% in the multi-camera pedestrian occupancy task, the DIC method costs around 130 KB in communication overhead.
In comparison, the communication cost of our TOCOM-TEM method with parameter $\tau_{2} = 0$ and $\tau_{2} = 1$ are 50 KB and 20 KB, respectively.
Fig. \ref{fig:WILDTRACK_entropy_model_visualization_example} shows visualizations of the normalized bitrate allocations for the input frames.
The second row of the figure corresponds to the extracted features of the DIC method.
It appears that its entropy model allocates many bits to some unique areas in an image to preserve all its characteristics in the reconstructed image.
However, this data-oriented communication scheme leads to high communication overhead, and a large part of the information is irrelevant to the downstream task.
On the contrary, the features of our method shown in the third and fourth rows mainly focus on task-relevant areas (e.g., the foot points of pedestrians) that are useful to predict occupancy.
This is because the task-oriented communication framework is optimized for a specific task, rather than for reliable video transmission.
Besides, setting $\tau_{2} = 1$ allows the temporal entropy model to utilize the previous feature as side information in entropy coding, which further reduces the communication overhead.

\section{Conclusions}
\label{Sec:Conclusion}


This study demonstrated the effectiveness of task-oriented communication for edge video analytics, which serves for video analytics applications rather than for video reconstruction.
Our proposed communication framework is built upon the deterministic IB principle and the temporal entropy model for feature extraction and encoding, respectively.
Besides, a spatial-temporal fusion module was proposed to integrate both the current received features and the previous features at the server to improve the performance of the edge video analytics system.
Compared with the data-oriented communication strategies, our proposed TOCOM-TEM method consistently outperformed the baselines in terms of the rate-performance tradeoff.

For future research, it is interesting to further exploit the spatial-temporal correlation of multiple devices in edge video analytics systems, and formalize the optimal rate-performance tradeoff.
This would have a broad impact on the intelligent video analytics applications that necessitate cooperation among multiple cameras.

\appendices

\section{Derivation of the upper bound of (\ref{equ:multi_device_extraction})}

\label{appendix:loss_1}

Recall that the objective function in (\ref{equ:multi_device_extraction}) includes the term $-I(Y_{t},\ldots, Y_{t+\tau_{1}} ; \hat{Z}_{t}^{(1: K)}) =  
H(Y_{t},\ldots,Y_{t+\tau_{1}}|\hat{Z}_{t}^{(1: K)}) - H(Y_{t},\ldots,Y_{t+\tau_{1}})  $ .
Writing the conditional entropy with the variational distribution, it becomes:
\begin{equation*}
\begin{aligned}
    & H(Y_{t},\ldots,Y_{t+\tau_{1}}|\hat{Z}_{t}^{(1: K)}) \\
    & = \mathbb{E}\left\{ -  \log   q(\bm{y}_{y},\ldots,\bm{y}_{t+\tau_{1}}|\hat{\bm{z}}_{t}^{(1:K)};\bm{\varphi}_{0:\tau_{1}}) \right\} \\
     & \quad - \underbrace{\mathbb{E}\left\{ \log \frac{p(\bm{y}_{y},\ldots,\bm{y}_{t+\tau_{1}}|\hat{\bm{z}}_{t}^{(1:K)})}{q(\bm{y}_{y},\ldots,\bm{y}_{t+\tau_{1}}|\hat{\bm{z}}_{t}^{(1:K)};\bm{\varphi}_{0:\tau_{1}})} \right\}}_{D_{\mathrm{KL}}(p\|q)\geq 0}.
\end{aligned}
\end{equation*}
As the KL-divergence is not negative, we have
\begin{equation}
\begin{aligned}
\label{equ:appendix_proof_1}
 & H(Y_{t},\ldots,Y_{t+\tau_{1}}|\hat{Z}_{t}^{(1: K)}) \\
 & \leq  \mathbb{E}\left\{ -\log   q(\bm{y}_{y},\ldots,\bm{y}_{t+\tau_{1}}|\hat{\bm{z}}_{t}^{(1:K)}) \right\}.
 \end{aligned}
\end{equation}
As the joint entropy $H(Y_{t},\ldots,Y_{t+\tau_{1}})$ is a constant, minimizing $\mathbb{E}\{ -\log   q(\bm{y}_{y},\ldots,\bm{y}_{t+\tau_{1}}|\hat{\bm{z}}_{t}^{(1:K)}) \}$ is equivalent to minimizing an upper bound of $ -  I(Y_{t},\ldots, Y_{t+\tau_{1}};\hat{Z}_{t}^{(1:K)})$. 

The entropy $H(\hat{Z}_{t}^{(k)})$ in (\ref{equ:multi_device_extraction}) represents the communication cost.
In the proposed method, we introduce a latent variable $\hat{\bm{v}}_{t}^{(k)}$ as side information to encode $\hat{z}_{t}^{(k)}$.
Both the quantized feature $\hat{\bm{z}}_{t}^{(k)}$ and the latent $\hat{\bm{v}}_{t}^{(k)}$ should be transmitted from the device to the server.
Therefore, the minimum communication cost is the joint entropy $H(\hat{\bm{z}}_{t}^{(k)},\hat{\bm{v}}_{t}^{(k)}) \geq H(\hat{\bm{z}}_{t}^{(k)})$.
Write the joint entropy term out in full with the variational distributions $q(\hat{\bm{z}}_t^{(k)} | \hat{\bm{v}}_t^{(k)} ; \bm{\psi}_d^{(k)})$ and $q(\hat{\bm{v}}_t^{(k)} ; \bm{\psi}_f^{(k)})$.
It is as follows:
\begin{equation*}
\begin{aligned}
H(\hat{\bm{z}}_{t}^{(k)},\hat{\bm{v}}_{t}^{(k)}) 
= & \mathbb{E} \left\{ -\log q(\hat{\bm{z}}_t^{(k)} | \hat{\bm{v}}_t^{(k)} ; \bm{\psi}_d^{(k)}) q(\hat{\bm{v}}_t^{(k)} ; \bm{\psi}_f^{(k)}) \right\} \\
& - \underbrace{\mathbb{E} \left\{ \log \frac{p(\hat{\bm{z}}_{t}^{(k)},\hat{\bm{v}}_{t}^{(k)})}{q(\hat{\bm{z}}_t^{(k)} | \hat{\bm{v}}_t^{(k)} ; \bm{\psi}_d^{(k)})q(\hat{\bm{v}}_t^{(k)} ; \bm{\psi}_f^{(k)})} \right\}}_{D_{\mathrm{KL}}(p\|q) \geq 0}.
\end{aligned}
\end{equation*}
As the KL-divergence is larger than zero, we have
\begin{equation}
\begin{aligned}
\label{equ:appendix_proof_2}
H(\hat{\bm{z}}_{t}^{(k)},\hat{\bm{v}}_{t}^{(k)}) \leq \mathbb{E} \left\{ -\log q(\hat{\bm{z}}_t^{(k)} | \hat{\bm{v}}_t^{(k)} ; \bm{\psi}_d^{(k)}) q(\hat{\bm{v}}_t^{(k)} ; \bm{\psi}_f^{(k)}) \right\}.
\end{aligned}
\end{equation}
By combining (\ref{equ:appendix_proof_1}) and (\ref{equ:appendix_proof_2}), we can find that minimizing the $\mathcal{L}_{1}$ loss in (\ref{equ:loss_feature_extraction}) is equivalent to minimizing an upper bound of (\ref{equ:multi_device_extraction}).

\appendices

\bibliographystyle{IEEEtran}
\bibliography{IEEEabrv,ref}

\begin{thebibliography}{10}
\providecommand{\url}[1]{#1}
\csname url@samestyle\endcsname
\providecommand{\newblock}{\relax}
\providecommand{\bibinfo}[2]{#2}
\providecommand{\BIBentrySTDinterwordspacing}{\spaceskip=0pt\relax}
\providecommand{\BIBentryALTinterwordstretchfactor}{4}
\providecommand{\BIBentryALTinterwordspacing}{\spaceskip=\fontdimen2\font plus
\BIBentryALTinterwordstretchfactor\fontdimen3\font minus \fontdimen4\font\relax}
\providecommand{\BIBforeignlanguage}[2]{{%
\expandafter\ifx\csname l@#1\endcsname\relax
\typeout{** WARNING: IEEEtran.bst: No hyphenation pattern has been}%
\typeout{** loaded for the language `#1'. Using the pattern for}%
\typeout{** the default language instead.}%
\else
\language=\csname l@#1\endcsname
\fi
#2}}
\providecommand{\BIBdecl}{\relax}
\BIBdecl

\bibitem{chang2020ai_smart_transportation}
M.-C. Chang, C.-K. Chiang, C.-M. Tsai, Y.-K. Chang, H.-L. Chiang, Y.-A. Wang, S.-Y. Chang, Y.-L. Li, M.-S. Tsai, and H.-Y. Tseng, ``A{I} city challenge 2020-computer vision for smart transportation applications,'' in \emph{Proc. Conf. Comput. Vision Pattern Recognit. Workshop}, Jun. 2020. [Online]. Available: https://ieeexplore.ieee.org/document/9151055.

\bibitem{othman2017new_surveillance}
N.~A. Othman and I.~Aydin, ``A new iot combined body detection of people by using computer vision for security application,'' in \emph{Proc. Int. Conf. Comput. Intell. Commun. Netw.}, Ercan, North Cyprus, Sep. 2017.

\bibitem{gao2018computer_healthcare}
J.~Gao, Y.~Yang, P.~Lin, and D.~S. Park, ``Computer vision in healthcare applications,'' \emph{J. Healthcare Eng.}, vol. 2018, Mar. 2018.

\bibitem{ananthanarayanan2017real_video_analytics}
G.~Ananthanarayanan, P.~Bahl, P.~Bod{\'\i}k, K.~Chintalapudi, M.~Philipose, L.~Ravindranath, and S.~Sinha, ``Real-time video analytics: The killer app for edge computing,'' \emph{Comput.}, vol.~50, no.~10, pp. 58--67, Oct. 2017.

\bibitem{zhou2019edge_edge_intell}
Z.~Zhou, X.~Chen, E.~Li, L.~Zeng, K.~Luo, and J.~Zhang, ``Edge intelligence: Paving the last mile of artificial intelligence with edge computing,'' \emph{Proc. IEEE}, vol. 107, no.~8, pp. 1738--1762, Aug. 2019.

\bibitem{ShaoTWC}
J.~Shao, Y.~Mao, and J.~Zhang, ``Task-oriented communication for multi-device cooperative edge inference,'' \emph{IEEE Trans. Wireless Commun.}, Jul. 2022.

\bibitem{shao2021learning}
J.~\vspace{0mm}Shao, Y.~Mao, and J.~Zhang, ``Learning task-oriented communication for edge inference: An information bottleneck approach,'' \emph{IEEE J. Sel. Area Commun.}, vol.~40, no.~1, pp. 197--211, Jan. 2022.

\bibitem{hanna2020distributed}
O.~A. Hanna, Y.~H. Ezzeldin, T.~Sadjadpour, C.~Fragouli, and S.~Diggavi, ``On distributed quantization for classification,'' \emph{IEEE J. Sel. Areas Inf. Theory}, vol.~1, no.~1, pp. 237--249, Apr. 2020.

\bibitem{singhal2020communication}
M.~Singhal, V.~Raghunathan, and A.~Raghunathan, ``Communication-efficient view-pooling for distributed multi-view neural networks,'' in \emph{Proc. Design Automat. Test Eur. Conf. Exhib. (DATE)}, Grenoble, France, Sep. 2020.

\bibitem{choi2019context}
J.~Choi, Z.~Hakimi, P.~W. Shin, J.~Sampson, and V.~Narayanan, ``Context-aware convolutional neural network over distributed system in collaborative computing,'' in \emph{Proc. ACM/IEEE Design Automat. Conf. (DAC)}, Las Vegas, NV, USA, Jun. 2019.

\bibitem{Multi-Robot_Collaborative_Percep_GNN}
Y.~Zhou, J.~Xiao, Y.~Zhou, and G.~Loianno, ``Multi-robot collaborative perception with graph neural networks,'' \emph{IEEE Robot. Automat. Lett.}, Jan. 2022.

\bibitem{xie2022robust_IB}
S.~Xie, Y.~Wu, S.~Ma, M.~Ding, Y.~Shi, and M.~Tang, ``Robust information bottleneck for task-oriented communication with digital modulation,'' \emph{arXiv preprint arXiv:2209.10382}, 2022. [Online]. Available: https://arxiv.org/abs/2209.10382.

\bibitem{tishby2000informationIB}
N.~Tishby, F.~C. Pereira, and W.~Bialek, ``The information bottleneck method,'' in \emph{Proc. Annu. Allerton Conf. Commun. Control Comput.}, Monticello, IL, USA, Oct. 2000.

\bibitem{sundermeyer2012lstm_LSTM}
M.~Sundermeyer, R.~Schl{\"u}ter, and H.~Ney, ``L{S}{T}{M} neural networks for language modeling,'' in \emph{Annu. Conf. of Int. Speech Commun. Assoc.}, Portland, OR, USA., Sep. 2012.

\bibitem{vaswani2017attention}
A.~Vaswani, N.~Shazeer, N.~Parmar, J.~Uszkoreit, L.~Jones, A.~N. Gomez, {\L}.~Kaiser, and I.~Polosukhin, ``Attention is all you need,'' \emph{Adv. Neural Inf. Process. Syst.}, vol.~30, Dec. 2017.

\bibitem{ResNet}
K.~He, X.~Zhang, S.~Ren, and J.~Sun, ``Deep residual learning for image recognition,'' in \emph{Proc. IEEE {C}onf. {C}omput. {V}is. {P}attern {R}ecognit.}, Las Vegas, NV, USA, Jun. 2016, pp. 770--778.

\bibitem{ristani2018features_pedestrian_tracking}
E.~Ristani and C.~Tomasi, ``Features for multi-target multi-camera tracking and re-identification,'' in \emph{Proc. Conf. Comput. Vision Pattern Recognit.}, Salt Lake City, UT, USA, Jun. 2018, pp. 6036--6046.

\bibitem{liu2016deep_vehicle_reid}
H.~Liu, Y.~Tian, Y.~Yang, L.~Pang, and T.~Huang, ``Deep relative distance learning: Tell the difference between similar vehicles,'' in \emph{Proc. Conf. Computer Vision Pattern Recognit.}, Las Vegas, NV, USA, Jun. 2016, pp. 2167--2175.

\bibitem{farneback2003two_motion_estimation}
G.~Farneb{\"a}ck, ``Two-frame motion estimation based on polynomial expansion,'' in \emph{Scandinavian Conf. Image Anal.}, Halmstad, Sweden, Jun.-Jul. 2003, pp. 363--370.

\bibitem{deterministic_ib}
D.~Strouse and D.~J. Schwab, ``The deterministic information bottleneck,'' \emph{Neural Comput.}, vol.~29, no.~6, pp. 1611--1630, Jun. 2017.

\bibitem{strinati20216g_goal_oriented}
E.~C. Strinati and S.~Barbarossa, ``6{G} networks: {B}eyond shannon towards semantic and goal-oriented communications,'' \emph{Comput. Netw,}, vol. 190, p. 107930, May 2021.

\bibitem{zhu2020toward}
G.~Zhu, D.~Liu, Y.~Du, C.~You, J.~Zhang, and K.~Huang, ``Toward an intelligent edge: wireless communication meets machine learning,'' \emph{IEEE Commun. Mag.}, vol.~58, no.~1, pp. 19--25, Jan. 2020.

\bibitem{shao2020communication}
J.~Shao and J.~Zhang, ``Communication-computation trade-off in resource-constrained edge inference,'' \emph{IEEE Commun. Mag.}, vol.~58, no.~12, pp. 20--26, Dec. 2020.

\bibitem{pappas2021goal}
N.~Pappas and M.~Kountouris, ``Goal-oriented communication for real-time tracking in autonomous systems,'' in \emph{2021 IEEE Int. Conf. Auton. Systems (ICAS)}.\hskip 1em plus 0.5em minus 0.4em\relax Montreal, QC, Canada: IEEE, Aug. 2021, pp. 1--5.

\bibitem{shlezinger2021deep}
N.~Shlezinger and Y.~C. Eldar, ``Deep task-based quantization,'' \emph{Entropy}, vol.~23, no.~1, p. 104, Jan. 2021.

\bibitem{merluzzi2021wireless}
M.~Merluzzi, P.~Di~Lorenzo, and S.~Barbarossa, ``Wireless edge machine learning: Resource allocation and trade-offs,'' \emph{IEEE Access}, vol.~9, pp. 45\,377--45\,398, Mar. 2021.

\bibitem{xie2021deep_semantic_NLP}
H.~Xie, Z.~Qin, G.~Y. Li, and B.-H. Juang, ``Deep learning enabled semantic communication systems,'' \emph{IEEE Trans. Signal Process.}, vol.~69, pp. 2663--2675, Apr. 2021.

\bibitem{bourtsoulatze2019deep}
E.~Bourtsoulatze, D.~B. Kurka, and D.~G{\"u}nd{\"u}z, ``Deep joint source-channel coding for wireless image transmission,'' \emph{IEEE Trans. Cogn. Commun. Netw.}, vol.~5, no.~3, pp. 567--579, May 2019.

\bibitem{iccshao}
J.~Shao and J.~Zhang, ``Bottlenet++: An end-to-end approach for feature compression in device-edge co-inference systems,'' in \emph{Proc. IEEE Int. Conf. Commun. Workshop}, Dublin, Ireland, Jun. 2020.

\bibitem{kang2022task_oriented_ref1}
X.~Kang, B.~Song, J.~Guo, Z.~Qin, and F.~R. Yu, ``Task-oriented image transmission for scene classification in unmanned aerial systems,'' \emph{IEEE Trans. Commun.}, vol.~70, no.~8, pp. 5181--5192, Jun. 2022.

\bibitem{dubois2021lossy}
Y.~Dubois, B.~Bloem-Reddy, K.~Ullrich, and C.~J. Maddison, ``Lossy compression for lossless prediction,'' in \emph{Proc. Int. Conf. Learn. Represent. Workshop on Neural Compression}, May 2021. [Online]. Available: https://openreview.net/forum?id=GfCs5NhoR8Q.

\bibitem{li2023task_hongru}
H.~Li, W.~Yu, H.~He, J.~Shao, S.~Song, J.~Zhang, and K.~B. Letaief, ``Task-oriented communication with out-of-distribution detection: An information bottleneck framework,'' \emph{arXiv preprint arXiv:2305.12423}, 2023. [Online]. Available: https://arxiv.org/abs/2305.12423.

\bibitem{bottlenet}
A.~E. Eshratifar, A.~Esmaili, and M.~Pedram, ``Bottlenet: A deep learning architecture for intelligent mobile cloud computing services,'' 2019. [Online]. Available: https://arxiv.org/abs/1902.01000.

\bibitem{jankowski2020wireless_Jankowski}
M.~{Jankowski}, D.~{Gündüz}, and K.~{Mikolajczyk}, ``Wireless image retrieval at the edge,'' \emph{IEEE J. Sel. Areas Commun.}, vol.~39, no.~1, pp. 89--100, May 2021.

\bibitem{shao2020branchy}
J.~Shao, H.~Zhang, Y.~Mao, and J.~Zhang, ``Branchy-gnn: A device-edge co-inference framework for efficient point cloud processing,'' in \emph{Proc. IEEE Int. Conf. Acoust. Speech Signal Process. (ICASSP)}, Toronto, Canada, Jun. 2021.

\bibitem{wiegand2003overview_AVC}
T.~Wiegand, G.~J. Sullivan, G.~Bjontegaard, and A.~Luthra, ``Overview of the h. 264/{A}{V}{C} video coding standard,'' \emph{IEEE Trans. Circuits Syst. Video technol.}, vol.~13, no.~7, pp. 560--576, Jul. 2003.

\bibitem{sullivan2012overview_HECV}
G.~J. Sullivan, J.-R. Ohm, W.-J. Han, and T.~Wiegand, ``Overview of the high efficiency video coding ({H}{E}{V}{C}) standard,'' \emph{IEEE Trans. Circuits Syst. Video technol.}, vol.~22, no.~12, pp. 1649--1668, Dec. 2012.

\bibitem{lu2019_dvc}
G.~Lu, W.~Ouyang, D.~Xu, X.~Zhang, C.~Cai, and Z.~Gao, ``D{V}{C}: An end-to-end deep video compression framework,'' in \emph{Proc. Conf. Comput. Vision Pattern Recognit.}, Georgia Tech, GA, USA, Jun. 2019, pp. 11\,006--11\,015.

\bibitem{liu2020conditional_conditional_entropy_model}
J.~Liu, S.~Wang, W.-C. Ma, M.~Shah, R.~Hu, P.~Dhawan, and R.~Urtasun, ``Conditional entropy coding for efficient video compression,'' in \emph{Eur. Conf. Comput. Vision}, Aug. 2020. [Online]. Available: https://link.springer.com/chapter/10.1007/978-3-030-58520-4\_27.

\bibitem{tung2022deepwive}
T.-Y. Tung and D.~G{\"u}nd{\"u}z, ``Deepwive: Deep-learning-aided wireless video transmission,'' \emph{IEEE J. Sel. Areas Commun.}, vol.~40, no.~9, pp. 2570--2583, Jul. 2022.

\bibitem{liu2020deep}
D.~Liu, Y.~Li, J.~Lin, H.~Li, and F.~Wu, ``Deep learning-based video coding: A review and a case study,'' \emph{ACM Comput. Surv. (CSUR)}, vol.~53, no.~1, pp. 1--35, Feb. 2020.

\bibitem{ma2019image}
S.~Ma, X.~Zhang, C.~Jia, Z.~Zhao, S.~Wang, and S.~Wang, ``Image and video compression with neural networks: A review,'' \emph{IEEE Trans. Circuits Syst. Video Technol.}, vol.~30, no.~6, pp. 1683--1698, Apr. 2019.

\bibitem{agustsson2020scale}
E.~Agustsson, D.~Minnen, N.~Johnston, J.~Balle, S.~J. Hwang, and G.~Toderici, ``Scale-space flow for end-to-end optimized video compression,'' in \emph{Proc. Conf. Comput. Vision Pattern Recognit.}, Jun 2020. [Online]. Available: https://ieeexplore.ieee.org/document/9157366.

\bibitem{hu2021fvc}
Z.~Hu, G.~Lu, and D.~Xu, ``Fvc: A new framework towards deep video compression in feature space,'' in \emph{Proc. Conf. Comput. Vision Pattern Recognit.}, Jun 2021, pp. 1502--1511.

\bibitem{li2021deep}
J.~Li, B.~Li, and Y.~Lu, ``Deep contextual video compression,'' in \emph{Proc. Adv. Neural Inf. Process. Syst.}, Dec. 2021. [Online]. Available: https://openreview.net/forum?id=evqzNxmXsl3.

\bibitem{ristani2016performance_duke_dataset}
E.~Ristani, F.~Solera, R.~Zou, R.~Cucchiara, and C.~Tomasi, ``Performance measures and a data set for multi-target, multi-camera tracking,'' in \emph{Eur. Conf. Comput. Vision}, Amsterdam, The Netherlands, Oct. 2016, pp. 17--35.

\bibitem{hou2020multiview_multiview_detection}
Y.~Hou, L.~Zheng, and S.~Gould, ``Multiview detection with feature perspective transformation,'' in \emph{Eur. Conf. Comput. Vision}, Aug. 2020. [Online]. Available: https://www.ecva.net/papers/ eccv\_2020/papers\_ECCV/papers/123520001.pdf.

\bibitem{letaief2021edge}
K.~B. Letaief, Y.~Shi, J.~Lu, and J.~Lu, ``Edge artificial intelligence for 6{G}: Vision, enabling technologies, and applications,'' \emph{IEEE J. Sel. Areas Commun.}, vol.~40, no.~1, pp. 5--36, Nov. 2021.

\bibitem{balle2018variational1}
J.~Ball{\'e}, D.~Minnen, S.~Singh, S.~J. Hwang, and N.~Johnston, ``Variational image compression with a scale hyperprior,'' in \emph{Proc. Int. Conf. Learn. Represent.}, Vancouver, BC, Canada, 2018.

\bibitem{balle2017end-to-end}
J.~Ball{\'e}, V.~Laparra, and E.~Simoncelli, ``End-to-end optimized image compression,'' in \emph{Proc. Int. Conf. Learn. Represent.}, Toulon, France, Apr. 2017.

\bibitem{minnen2018jointballe}
D.~Minnen, J.~Ball{\'e}, and G.~D. Toderici, ``Joint autoregressive and hierarchical priors for learned image compression,'' in \emph{Proc. Neural Inf. Process. Syst.}, Montreal, QC, Canada, Dec. 2018.

\bibitem{wildtrack_dataset}
T.~Chavdarova, P.~Baqu{\'e}, S.~Bouquet, A.~Maksai, C.~Jose, T.~Bagautdinov, L.~Lettry, P.~Fua, L.~Van~Gool, and F.~Fleuret, ``Wildtrack: A multi-camera {H}{D} dataset for dense unscripted pedestrian detection,'' in \emph{Proc. Conf. Comput. Vision Pattern Recognit.}, Salt Lake City, UT, USA, Jun. 2018, pp. 5030--5039.

\bibitem{MVMC_dataset}
G.~Roig, X.~Boix, H.~B. Shitrit, and P.~Fua, ``Conditional random fields for multi-camera object detection,'' in \emph{Proc. Int. Conf. Comput. Vision}, Barcelona, Spain, Nov. 2011, pp. 563--570.

\bibitem{MODA_metric}
R.~Kasturi, D.~Goldgof, P.~Soundararajan, V.~Manohar, J.~Garofolo, R.~Bowers, M.~Boonstra, V.~Korzhova, and J.~Zhang, ``Framework for performance evaluation of face, text, and vehicle detection and tracking in video: Data, metrics, and protocol,'' \emph{IEEE Trans. Pattern Anal. Mach. Intell.}, vol.~31, no.~2, pp. 319--336, Mar. 2008.

\bibitem{lin2014microsoft_coco}
T.-Y. Lin, M.~Maire, S.~Belongie, J.~Hays, P.~Perona, D.~Ramanan, P.~Doll{\'a}r, and C.~L. Zitnick, ``Microsoft {C}{O}{C}{O}: Common objects in context,'' in \emph{Eur. Conf. Comput. Vision}, Zurich, Switzerland, Sep. 2014, pp. 740--755.

\bibitem{si2016research_webp}
Z.~Si and K.~Shen, ``Research on the webp image format,'' in \emph{Adv. Graph. Commun. Packag. Technol. Mater.}, Dec. 2015, pp. 271--277.

\bibitem{begaint2020compressai_package}
J.~B{\'e}gaint, F.~Racap{\'e}, S.~Feltman, and A.~Pushparaja, ``Compress{A}{I}: {A} pytorch library and evaluation platform for end-to-end compression research,'' \emph{arXiv preprint arXiv:2011.03029}, 2020. [Online]. Available: https://arxiv.org/abs/2011.03029.

\bibitem{richardson2011h_H264}
I.~E. Richardson, \emph{The H. 264 advanced video compression standard}.\hskip 1em plus 0.5em minus 0.4em\relax John Wiley \& Sons, 2011.

\bibitem{tian2019fcos}
Z.~Tian, C.~Shen, H.~Chen, and T.~He, ``Fcos: Fully convolutional one-stage object detection,'' in \emph{Proc. Int. Conf. Comput. Vision}, Oct.-Nov. 2019, pp. 9627--9636.

\bibitem{lin2017feature_FPN}
T.-Y. Lin, P.~Doll{\'a}r, R.~Girshick, K.~He, B.~Hariharan, and S.~Belongie, ``Feature pyramid networks for object detection,'' in \emph{Proc. Conf. Comput. Vision Pattern Recognit.}, Honolulu, HI, USA, Jul. 2017, pp. 2117--2125.

\end{thebibliography}

\end{document}